 \documentclass[final,1p,times]{elsarticle}

\usepackage{amssymb}
\usepackage{amsmath}
\usepackage{hyperref}
\hypersetup{hidelinks}
\usepackage{tikz}

\usetikzlibrary{shapes.geometric}
\usepackage{subcaption}
\usepackage{xcolor}
\usepackage{multirow}

\begin{document}

\begin{frontmatter}



\title{Applicability of machine learning in uncertainty quantification of turbulence models}


\author[inst1]{Marcel Matha}

\affiliation[inst1]{organization={German Aerospace Center (DLR), Institute of Propulsion Technology},
            addressline={Linder Höhe}, 
            postcode={51147 Cologne}, 
            country={Germany}}

\author[inst1]{Karsten Kucharczyk}

\begin{abstract}
The aim of this work is to apply and analyze machine learning methods for uncertainty quantification of turbulence models. In this work we investigate the classical and data-driven variants of the eigenspace perturbation method. This methodology is designed to estimate the uncertainties related to the shape of the modeled Reynolds stress tensor in the Navier-Stokes equations for \textit{\textbf{C}omputational \textbf{F}luid \textbf{D}ynamics} (CFD). The underlying methodology is extended by adding a data-driven, physics-constrained machine learning approach in order to predict local perturbations of the Reynolds stress tensor. Using separated two-dimensional flows, we investigate the generalization properties of the machine learning models and shed a light on impacts of applying a data-driven extension.
\end{abstract}


\end{frontmatter}


\section{Introduction}
\label{sec:introduction}
The objective of this white paper is to derive, apply, and discuss the eigenspace perturbation method, in its purely physics-based data-free and data-driven variants. The purpose of the perturbation method is to estimate the uncertainty based on the chosen turbulence model for CFD simulations. The presented work is mainly based on Kucharczyk \cite{kucharczyk} and Matha et al. \cite{matha2022, matha2022CF}.\\

For numerical simulations of turbulent flows in industrial applications, the RANS approach (\textit{\textbf{R}eynolds-\textbf{a}veraged \textbf{N}avier-\textbf{S}tokes}) has become commonplace during past decades. The basic idea of the RANS approach is to average flow variables (such as velocity and pressure) over time. The derivation of the underlying RANS equations requires certain modeling for certain terms, which inherently introduces uncertainty into the solutions. The component to be modeled is the so-called Reynolds stress tensor and modeling of this tensor is also referred to as turbulence modeling.

Over decades, multiple turbulence models have been developed, based on a variety of approaches. Each model has its strengths and weaknesses in terms of applicability, accuracy and computational complexity. Due to their simplicity and cost-effectiveness, linear eddy viscosity models are often used for simulations of turbomachinery. One of the central tasks of turbulence modeling is to determine the uncertainties that arise from the RANS approach \cite{duraisamy2019}. There are several approaches to account for these. The perturbation method of Emory et al. \cite{emory_diss, emory2013} does not aim for quantifying the uncertainty based on the parameters of a turbulence model, but determines the uncertainty with respect to the shape of the Reynolds stress tensor (non-parametric). Based on certain assumptions while deriving turbulence models, such as a linear eddy viscosity model, a particular shape of the Reynolds stress tensor is imposed, which is generally not the accurate shape.  Determining this structural uncertainty is identified by Zang et al. \cite{zang2002} as one of the major challenges in uncertainty quantification for turbulence models.

In this work, the perturbation method requires only three additional flow simulations, each of them is performed with a manipulated Reynolds stress tensor. The solutions of these three additional simulations cover a possible solution space, determining the uncertainty regarding the shape of the stress tensor. In order to run the perturbation method, a perturbation strength must be specified as an input in advance, defining the desired amount of manipulating the modeled Reynolds stress tensor. The larger the perturbation strength, the more uncertainty can be accounted for. The initial data-free perturbation method is very conservative due to the choice of the most pessimistic value for the perturbation magnitude. For this reason, Heyse et al. \cite{heyse2021} have tried to predict the perturbation strength by a machine learning model using training data from so-called scale-resolving simulations (\textit{\textbf{D}irect \textbf{N}umerical \textbf{S}imulation} (DNS) and \textit{\textbf{L}arge \textbf{E}ddy \textbf{S}imulation} (LES)). 

We use the data-free and the data-driven perturbation method, implemented in DLR’s CFD solver suite TRACE \cite{matha2022, matha2022CF}. The two-equation, linear eddy viscosity Menter SST $k-\omega$ turbulence model is selected as the baseline model for all conducted RANS simulations \cite{menter2003}. Therefore, the resulting uncertainty estimates are also related to this specific turbulence model. In addition, we focus on the machine learning framework used and assess the generalization properties of the selected machine learning models. We also try to find answers to general research questions concerning the underlying methodology, which we think are not given satisfactorily in literature.

\section{Uncertainty Quantification}
\label{sec:uq_approach}
\subsection{Data-Free Approach}
\label{sec:data-free}
Due to the formulation of a turbulence model, assumptions with respect to the functional form of the Reynolds stress tensor have to be made. As these assumptions are universally not valid, the determination of uncertainties has become part of turbulence modeling research as well. This process is also called \textit{\textbf{U}ncertainty \textbf{Q}uantification} (UQ).

According to Duraisamy et al. \cite{duraisamy2019}, these approaches can be divided into two classes, the parametric and the non-parametric approaches. The parametric approaches are devoted to the uncertainties arising from the pre-specified parameters in the turbulence modeling equations (mostly transport equations). However, in this paper we discuss the non-parametric approach suggested by Emory \cite{emory_diss}. Due to the Boussinesq assumption ($\tau_{ij} = -2 \mu_t \left(S_{ij} - \frac{1}{3}\frac{\partial u_k}{\partial x_k}\delta_{ij}\right) + \frac{2}{3} k \delta_{ij}$)
a particular shape is strictly imposed on the Reynolds stress tensor, which is preserved even if the modeling parameters are changed. Thus, the basic idea of the underlying method is to change the individual components of the Reynolds stress tensor to account for the uncertainty based on the shape of the Reynolds stress tensor.

\subsubsection{The perturbation method}
\label{sec:uq_perturbationMethod}

The entries of the Reynolds stress tensor can generally be expressed as $\tau_{ij} = k \cdot \left(a_{ij} + \frac{2}{3} \cdot \delta_{ij}\right) = k \cdot \left(v_{in}\Lambda_{n\ell}v_{\ell j} + \frac{2}{3} \cdot \delta_{ij} \right)$.
In order to estimate the uncertainty that arises due to the shape of the Reynolds stress tensor, additional flow simulations featuring modified Reynolds stress tensors have to be performed. The resulting enclosing range for certain \textit{\textbf{Q}uantity \textbf{o}f \textbf{I}nterest} (QoI) is intended to reveal the possible solution space and is referred to as the uncertainty band based on the Reynolds stress tensor.

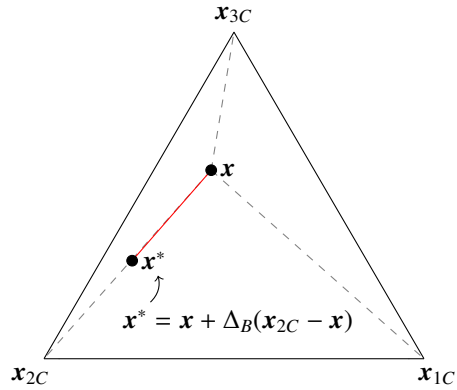
\begin{figure}[htb]
\centering
\begin{tikzpicture}
\draw [fill=white, white] (-3,-2) rectangle (3,2.9);
\draw [-](-2.5,-1.5) -- (2.5,-1.5) node[] {};
\draw [-](-2.5,-1.5) -- (0,2.83) node[] {};
\draw [-]( 2.5,-1.5) -- (0,2.83) node[] {};
\draw [gray, dashed]( -0.3, 1) -- (0,2.83) node[] {};
\draw [gray, dashed]( -0.3, 1) -- (-2.5,-1.5) node[] {};
\draw [gray, dashed]( -0.3, 1) -- (2.5,-1.5) node[] {};
\draw [red]( -1.34, -0.2) -- (-0.3, 1) node[] {};
\filldraw[black] (-0.3, 1) circle (2pt) node[anchor=west]{$\boldsymbol{x}$};
\filldraw[black] (-2.3, -1.5) circle (0.01pt) node[anchor=north east]{$\boldsymbol{x}_{2C}$};
\filldraw[black] (2.3, -1.5) circle (0.01pt) node[anchor=north west]{$\boldsymbol{x}_{1C}$};
\filldraw[black] (0, 2.83) circle (0.01pt) node[anchor=south]{$\boldsymbol{x}_{3C}$};
\filldraw[black] (-1.34, -0.2) circle (2pt) node[anchor=west]{$\boldsymbol{x}^*$};
\draw[->] (-1.1, -0.75) to[bend right] (-1.0, -0.4);
\node [] at (0.05, -1.0) {$\boldsymbol{x}^* = \boldsymbol{x} + \Delta_B(\boldsymbol{x}_{2C} - \boldsymbol{x})$};
\end{tikzpicture}
\caption[Schematic representation of the eigenvalue perturbation approach]{\small{Schematic representation of the eigenvalue perturbation approach towards $\boldsymbol{x}_{2C}.$}}
\label{img:perturbation_barycentric_sketch}
\end{figure}

Formally, the entries of the manipulated Reynolds stress tensor are given by
\begin{equation}
\label{gl:perturbatedReStress}
    \tau_{ij}^* = k^*\cdot \left(a_{ij}^* + \frac{2}{3} \cdot \delta_{ij}\right)= k^* \cdot \left(v_{in}^*\Lambda_{n\ell}^*v_{\ell j}^* + \frac{2}{3} \cdot \delta_{ij} \right) \ \text{.}
\end{equation}
The quantities marked with $*$ denote the perturbed components. Whereby the eigenvalues can be manipulated using a barycentric mapping, which is described by Banerjee et al. \cite{banerjee2007}. Emory et al. \cite{emory2013} proposed to consider the three vertices of the barycentric triangle of the Reynolds stress tensor as turbulence limiting states. The perturbed barycentric coordinates $\boldsymbol{x}^*$ are calculated based on the initial coordinate $\boldsymbol{x}$
\begin{equation}\label{gl:perturbation}
    \boldsymbol{x}^* = \boldsymbol{x} + \Delta_B \left(\boldsymbol{x}_{(t)} - \boldsymbol{x}\right) \ \text{,}
\end{equation}
where $\Delta_B \in [0,1]$ is the (relative) perturbation strength, and $\boldsymbol{x}_{\left(t\right)} \in \{\boldsymbol{x}_{1C}, \boldsymbol{x}_{2C}, \boldsymbol{x}_{3C}\}$ denotes a corner of the barycentric triangle (see Figure \ref{img:perturbation_barycentric_sketch}). The perturbed barycentric coordinates can be converted to an eigenvalue triple using the inverse of the barycentric mapping.

It is important to emphasize that the perturbation is controlled by the perturbation strength $\Delta_B$. Accordingly, if $\Delta_B=1$ is chosen, the barycentric coordinates in the entire computational domain are shifted to an identical limiting state (corresponding to the same location in the barycentric triangle). On the other hand, if $\Delta_B < 1$ is chosen, the barycentric coordinates for every computational grid point might be different.
The naive and consequently the most conservative approach is to choose $\Delta_B=1$ and determining the eigenvalues of the anisotropy as the respective limiting state. 

Unlike the eigenvalues, no direct constraints are known for the eigenvectors and the turbulent kinetic energy (except $k \geq 0$). Nevertheless, performing perturbation for the eigenvectors was introduced by Iaccarino et al. \cite{iaccarino2017} and also performed with TRACE by Matha et al. \cite{matha2022, matha2022CF}. 

Due to the nonlinearity of the Navier-Stokes equations, it is anything but obvious whether the three limiting states of turbulence are an appropriate choice as targeting points and what exactly is the impact of the perturbation strength $\Delta_B$. Since no precise answers to these questions could be found in literature, we consider pure eigenvalue perturbation (setting $V^*=V$ and $k^*=k$) to answer these research questions.

Besides, the eigenspace perturbation approach may lead to unstable simulations. To solve this issue, Mishra et al. \cite{mishra2018} suggested to use an additional moderation factor $f \in [0,1]$. For this purpose, a new Reynolds stress tensor $\tau_{{f}}^*$ is calculated, its entries can be written as
\begin{equation}
    (\tau_{f}^*)_{ij}= \tau_{ij} + f \cdot (\tau_{ij}^* - \tau_{ij}) \ \text{.}
\end{equation}
 The moderation factor handles the strength of the perturbation at the level of the Reynolds stress tensor, whereas the factor $\Delta_B$ describes the strength of the perturbation at the level of the barycentric coordinates. Nevertheless, it can be shown, that in the case of pure eigenvalue perturbation the effect of $f$ and $\Delta_B$ is identical \cite{kucharczyk, matha2022}. Instead of applying the moderating factor $f$ at the level of the Reynolds stress tensor, $\tau_{f}^*$ can be calculated directly by multiplying the perturbation strength $\Delta_B$ by $f$.

\subsubsection[Applying the data-free perturbation method]{Applying the data-free perturbation method}\label{section:4.2}

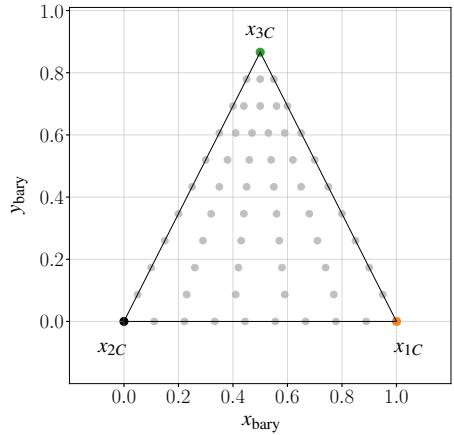
\begin{figure}[!htb]
\centering
\begin{tikzpicture}
\node[anchor=south west] (image) at (0,0) {
     \includegraphics[width=0.48\textwidth]{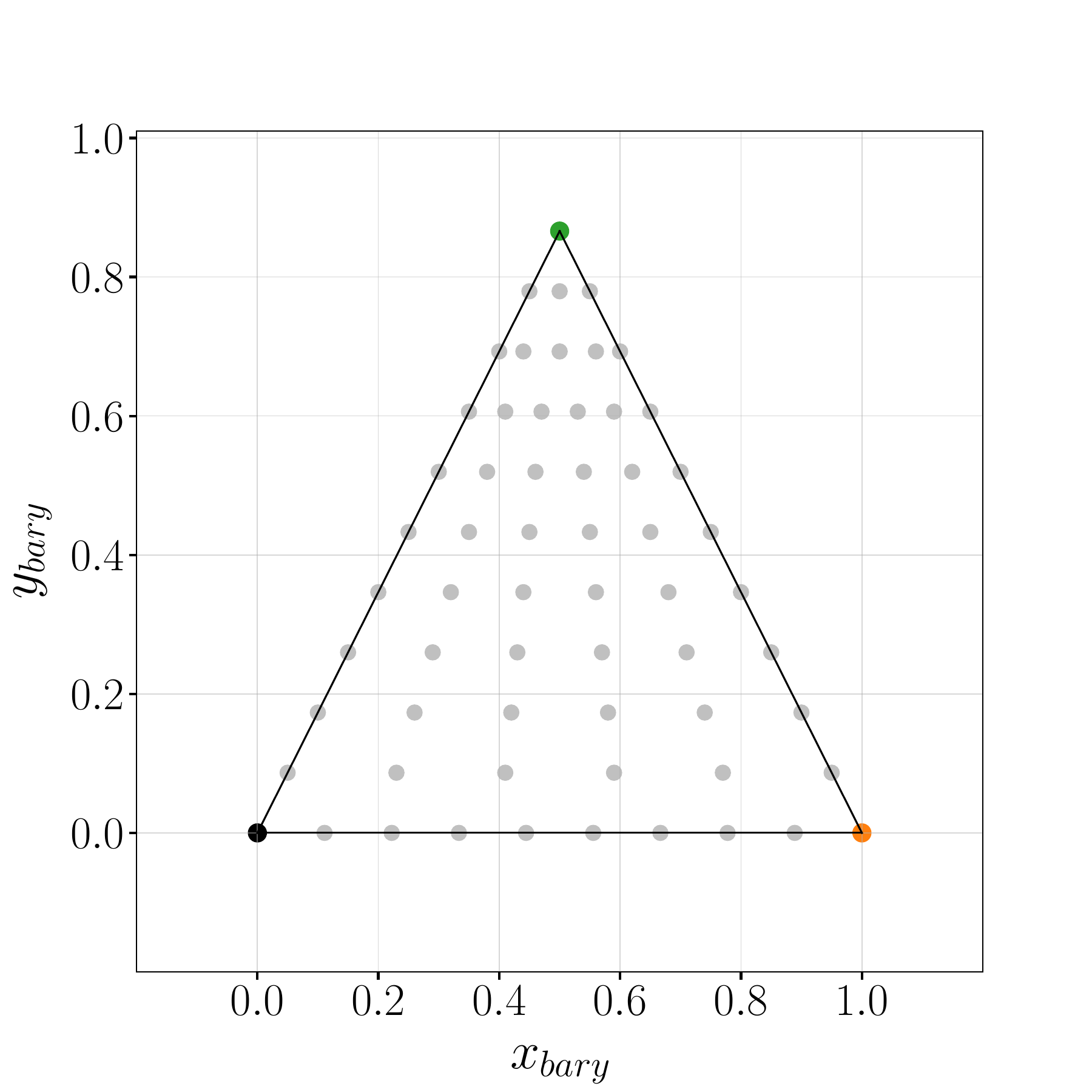}
    };
\node[] at (1.5,1.3) {\footnotesize{$x_{2C}$}};
\node[] at (5.4,1.3) {\footnotesize{$x_{1C}$}};
\node[] at (3.45,5.5) {\footnotesize{$x_{3C}$}};
\draw [fill=white, white] (2.0,0.0) rectangle (7.0,0.5);
\draw [fill=white, white] (0.1,0.0) rectangle (0.5,6.0);
    
\node[] at (3.46,0.3) {\footnotesize{$x_{\mathrm{bary}}$}};
\node[rotate=90] at (0.275,3.35) {\footnotesize{$y_{\mathrm{bary}}$}};
\end{tikzpicture}
\caption[Distribution of chosen barycentric coordinates]{\small{Distribution of chosen barycentric coordinates.}}
\label{fig:datenwolke_triangle}
\end{figure}

In this section, we apply the perturbation method to the converging-diverging channel simulations at $Re_\tau$ = 617. The simulation performed with TRACE can be compared to the DNS results by Laval and Marquillie in \cite{laval2011}.
In order to investigate how the solution space behaves when targeting different barycentric coordinates, the entire barycentric triangle is discretized with 61 points. Each point inside the barycentric triangle in Figure \ref{fig:datenwolke_triangle} corresponds to a perturbed target state simulation. The moderation factor $f$ is not considered here, which results in some solutions, which are not convergent. We discuss the issue of convergence in the discussion section of these results. 

\begin{figure}[htb]
\centering
   \begin{minipage}[c]{0.495\textwidth}
    \centering
    \includegraphics[width=\textwidth, height=8cm, trim=0.5cm 0.2cm 3cm 2cm, clip=True]{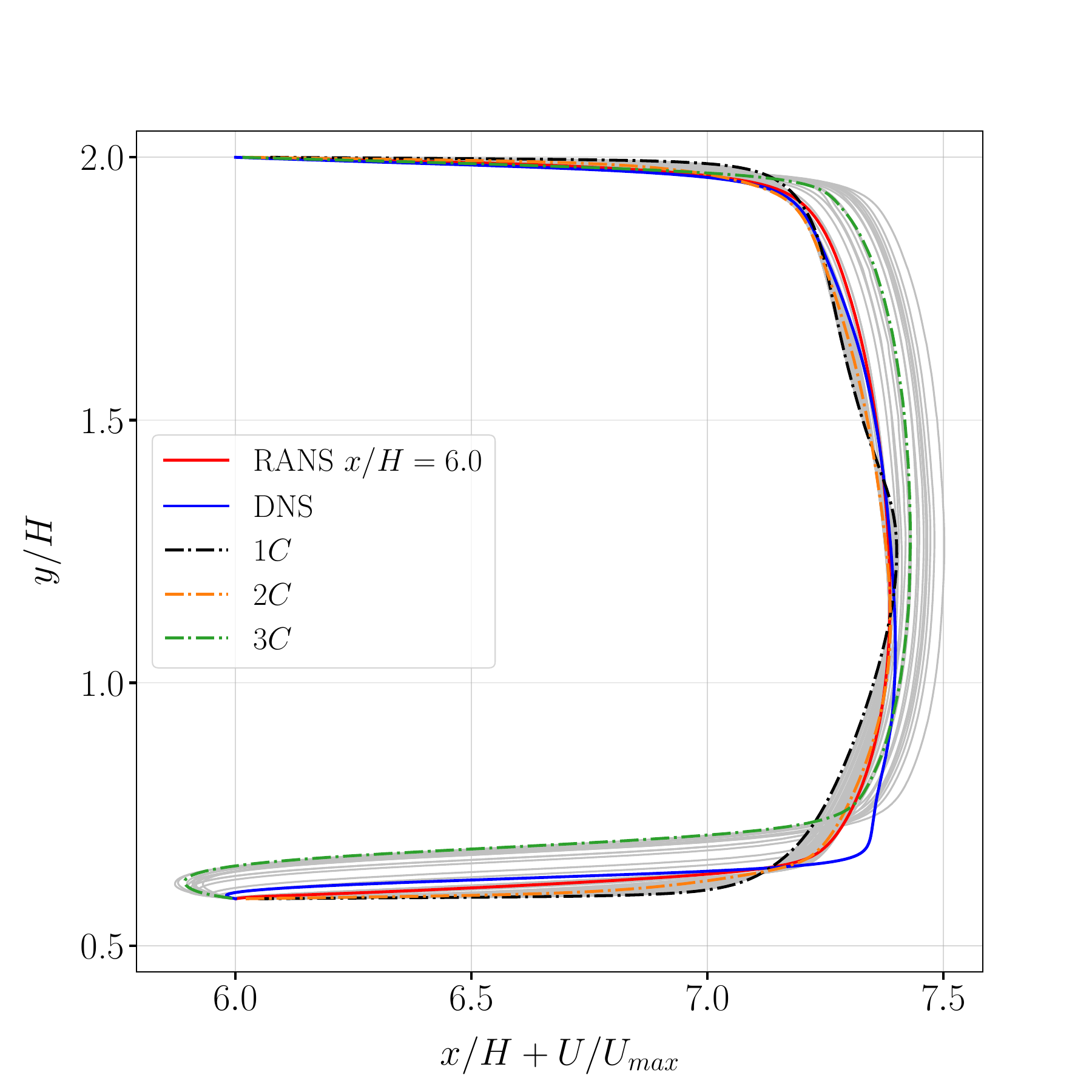}
    \end{minipage}
    \begin{minipage}[c]{0.495\textwidth}
    \centering
    \includegraphics[width=\textwidth, height=8cm, trim=0.5cm 0.2cm 3cm 2cm, clip=True]{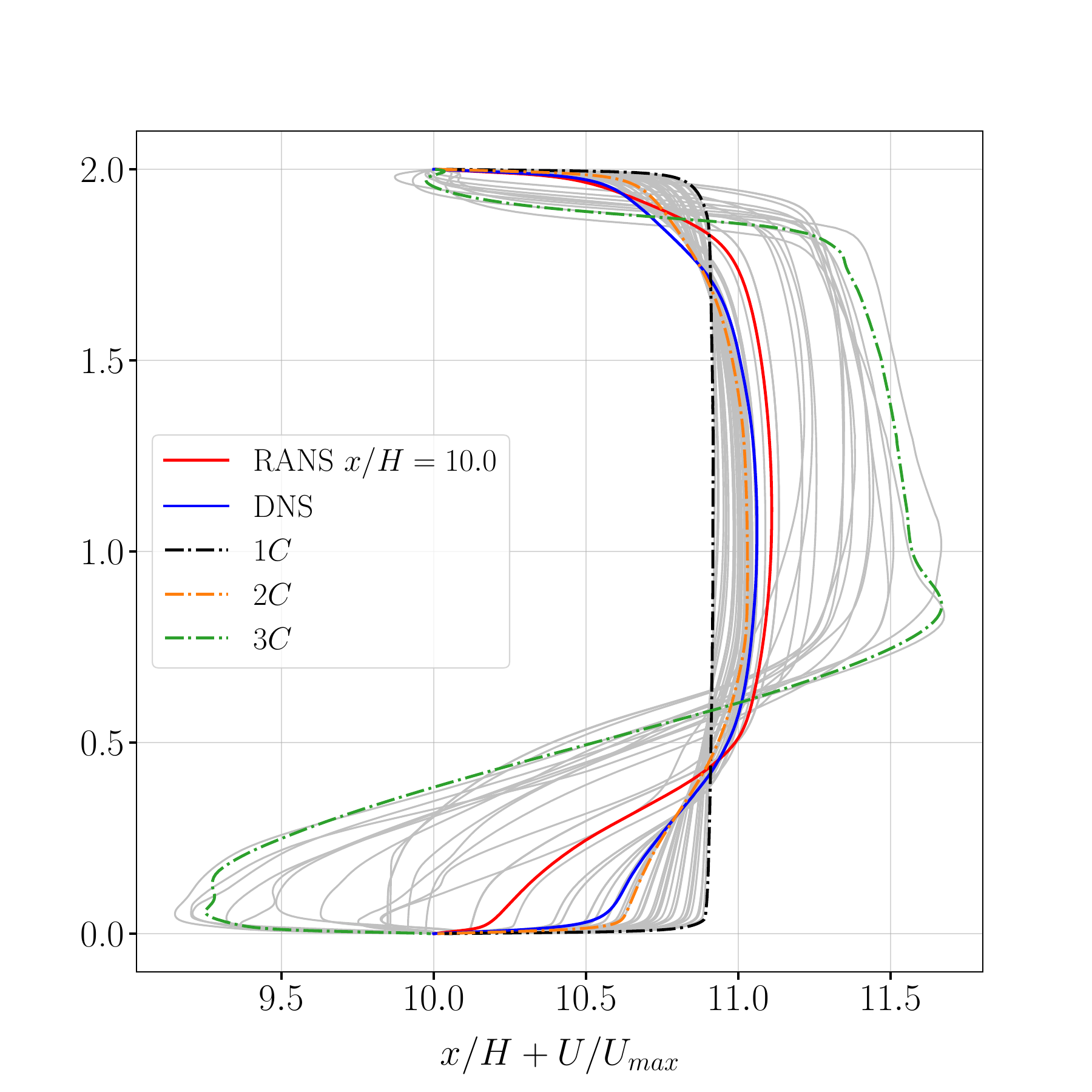}
    \end{minipage}
    \caption[Velocity profiles of the perturbed simulations]{\small{Resulting velocity profiles of the perturbed simulations based on the chosen barycentric coordinates in Figure \ref{fig:datenwolke_triangle} at $x/H=6.0$ and $x/H=10.0$.}}
   
    \label{fig:datenwolke_conv_div_}
\end{figure}

The resulting velocity profiles at $x/H =6.0$ and $x/H=10$ are presented in Figure \ref{fig:datenwolke_conv_div_}. First of all, it can be observed that the solutions of the $1C$ and $2C$ perturbations differ significantly from the solution of the $3C$ perturbation. The $3C$ perturbation increases the separated area compared to the baseline result, while the separation is suppressed for the other two perturbations. It can be seen that the solution of the $1C$ perturbation bounds the velocity profiles almost in the entire simulation domain to one side. Nevertheless, there are also gray colored velocity profiles, corresponding to data points inside the barycentric triangle (see Figure \ref{fig:datenwolke_triangle}), which are not bounded by the $3C$-perturbation and are the extreme states in case of the velocity profile consequently. This may be due to the fact, that by approaching the isotropic corner for the anisotropy tensor, the simulations tend to get unstable. Thus, it is difficult to answer the question whether the limiting states of turbulence lead to strictly bounding extreme values for certain QoI. We could have fixed the convergence issue by applying the moderation factor, but comparability with respect to the exact location inside the barycentric triangle would no longer be guaranteed. Nevertheless, as it can be seen based on the $1C$ perturbation, which is indeed an extreme condition for the velocity profile, we agree to conduct only three additional simulations targeting for the corners of the barycentric triangle to obtain a meaningful estimation of the uncertainty of the underlying turbulence model.  

\begin{figure}[!htb]
\centering
   \begin{minipage}[c]{0.495\textwidth}
    \centering
    \includegraphics[width=\textwidth, height=8cm, trim=0.5cm 0.2cm 3cm 2cm, clip=True]{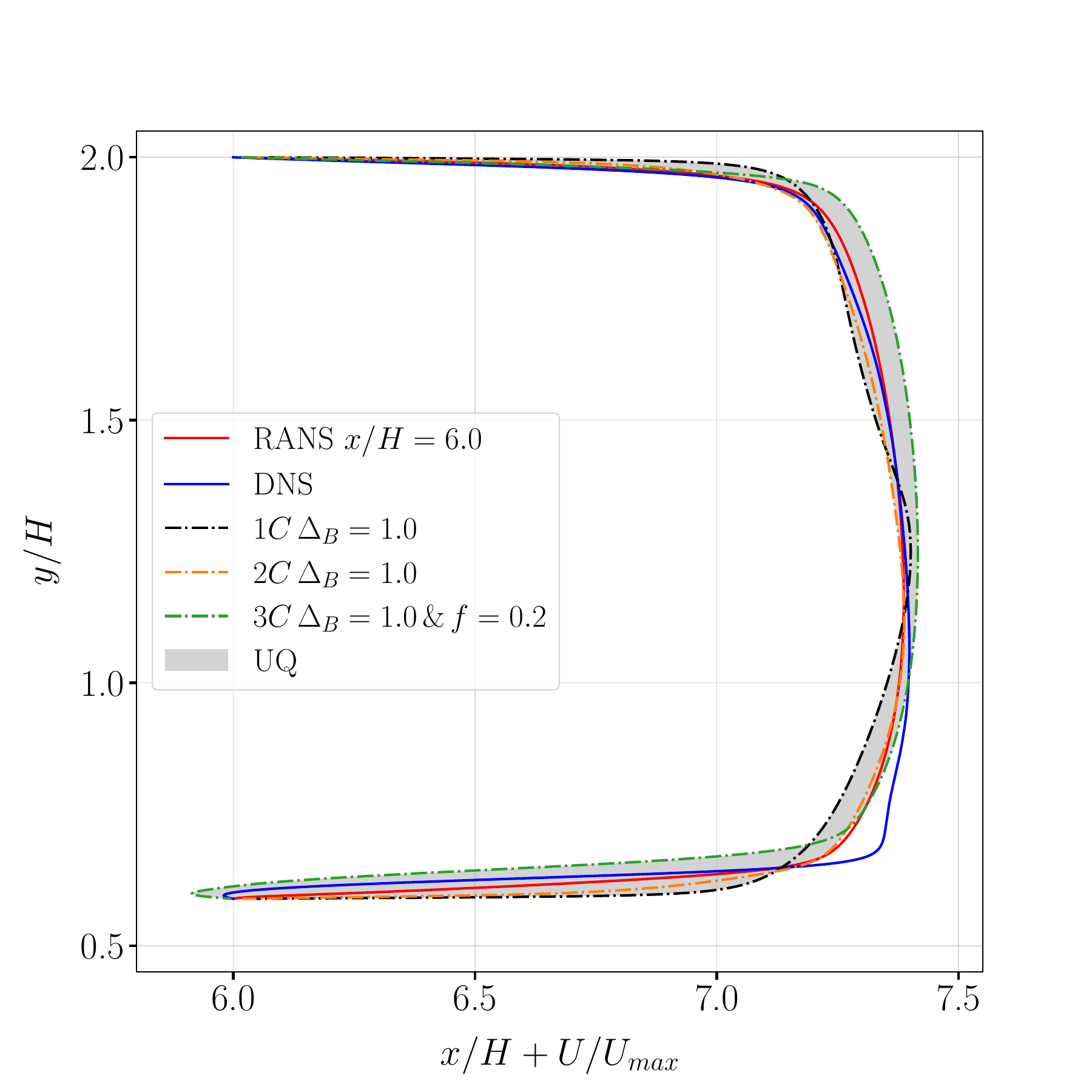}
    \end{minipage}
    \begin{minipage}[c]{0.495\textwidth}
    \centering
    \includegraphics[width=\textwidth, height=8cm, trim=0.5cm 0.2cm 3cm 2cm, clip=True]{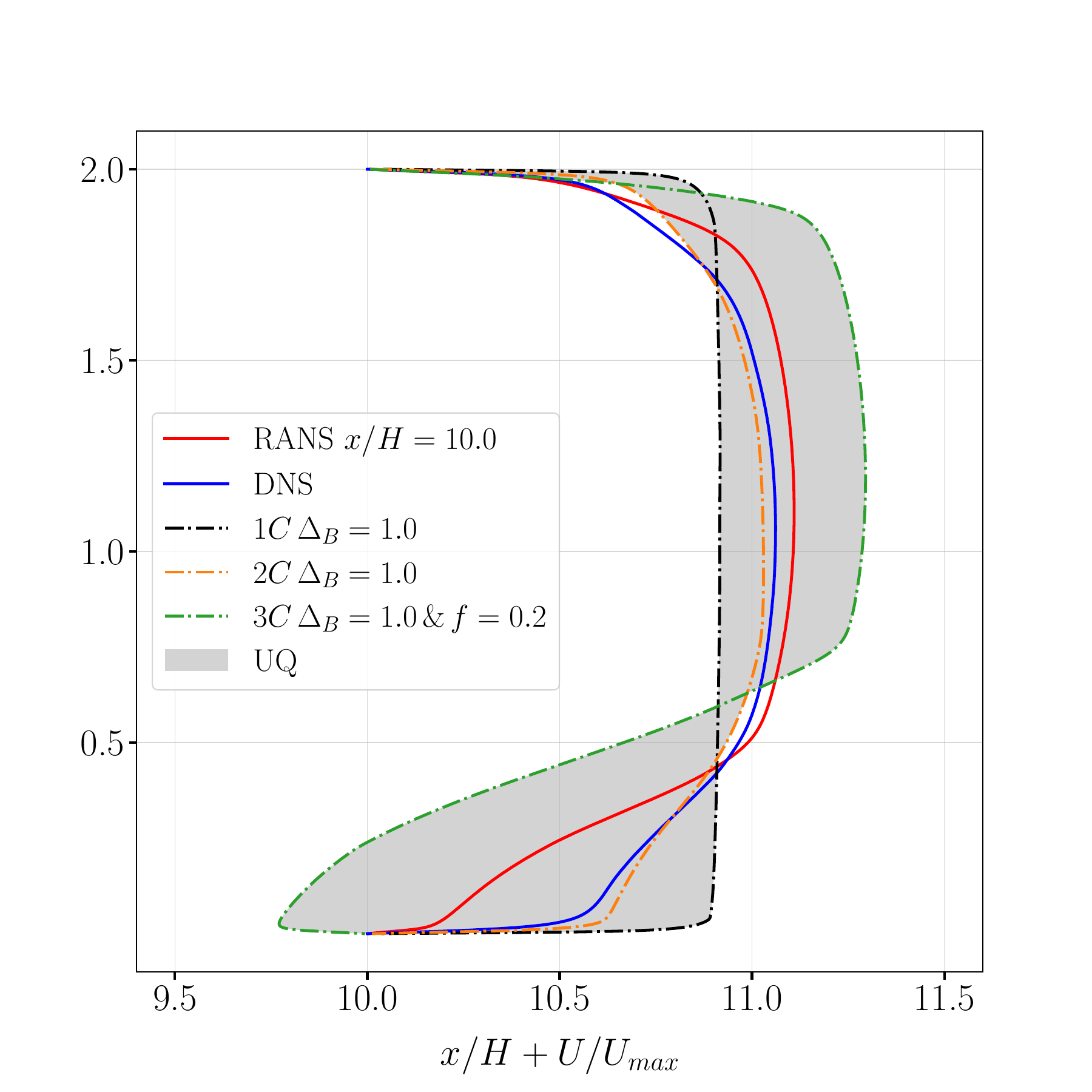}
    \end{minipage}
    \caption[Estimated uncertainty bands for the converging-diverging channel]{\small{Estimated uncertainty bands for the converging-diverging channel at $x/H=6.0$ and $x/H=10.0$.}}
   
    \label{fig:uq_band_final}
\end{figure}

Figure \ref{fig:uq_band_final} shows the final uncertainty band generated by the perturbation method, whereby a convergent solution for the $3C$ perturbation was achieved by choosing a moderation factor of $f=0.2$. 
From our point of view, it is not possible to qualitatively evaluate the correctness of the uncertainty band. Accordingly, the uncertainty band only indicates a range of possible solutions, which can be estimated by modifications of the turbulence model. In literature, such uncertainty bands are often evaluated using DNS or LES solutions. It can be seen that great part of the DNS solution is included in the uncertainty band. However, regions which do not cover the DNS solution can also be observed. One reason for this is that there is an infinite number of barycentric coordinates to manipulate the Reynolds stress tensor, which can not be capture adequately in reasonable computational time. Furthermore, only eigenvalue perturbation was considered, while keeping the eigenvectors and the turbulent kinetic energy constant throughout each perturbation. 
Moreover, there are also regions, where the unperturbed (baseline) RANS solution is not fully included in the uncertainty band. We shed a light on this after the next investigation. 

In practice, the uncertainty bounds have to be as conservative as needed. That is the reason why Heyse et al. \cite{heyse2021} propose a strategy to optimize the uncertainty bands. The basic idea is to determine a appropriate perturbation magnitude $\Delta_B$ in advance. The indicator of perturbation strength is determined using scale-resolving data and machine learning methods. Before we introducing this methodology, we investigate the effect of $\Delta_B$ in more detail.

Therefore, we evaluate the size of the uncertainty band for the velocity profile at $x/H=10.0$ with respect to the change in $\Delta_B \in \{0.1, 0.2, 0.5, 0.8, 1.0\}$. As we have already observed that the \mbox{$3C$-perturbation} does not provide a convergent solution without a moderation factor, a moderation factor of $f=0.2$ is used for each \mbox{$3C$-perturbation}. The resulting enclosing areas between the limiting QoI curves are summarized in Table \ref{tab:flaechen_delta_b}.

\begin{table}[!htb]
    \centering
    \footnotesize
    \begin{tabular}{ |c|c| } 
\hline
$\Delta_B$ & Area of the uncertainty band at $x/H=10.0$ \\
\hline
 0.1   & 0.189    \\
 \hline
 0.2   & 0.280   \\ 
 \hline
0.5   & 0.534   \\ 
\hline
0.8   & 0.780   \\
\hline
1.0   & 0.910   \\
\hline
\end{tabular}
    \caption[Are of the enclosed region of the uncertainty band for the velocity profile at $x/H=10.0$ with respect to $\Delta_B$]{\small{Size of the uncertainty band for the velocity profile at $x/H=10.0$ with respect to $\Delta_B$.}}
    \label{tab:flaechen_delta_b}
\end{table}

\begin{figure}[!htb]
\centering
\begin{tikzpicture}
\node[anchor=south west] (image) at (0,0) {
     \includegraphics[width=0.495\textwidth, height=8cm, trim=0.4cm 0.2cm 3cm 2cm, clip=True]{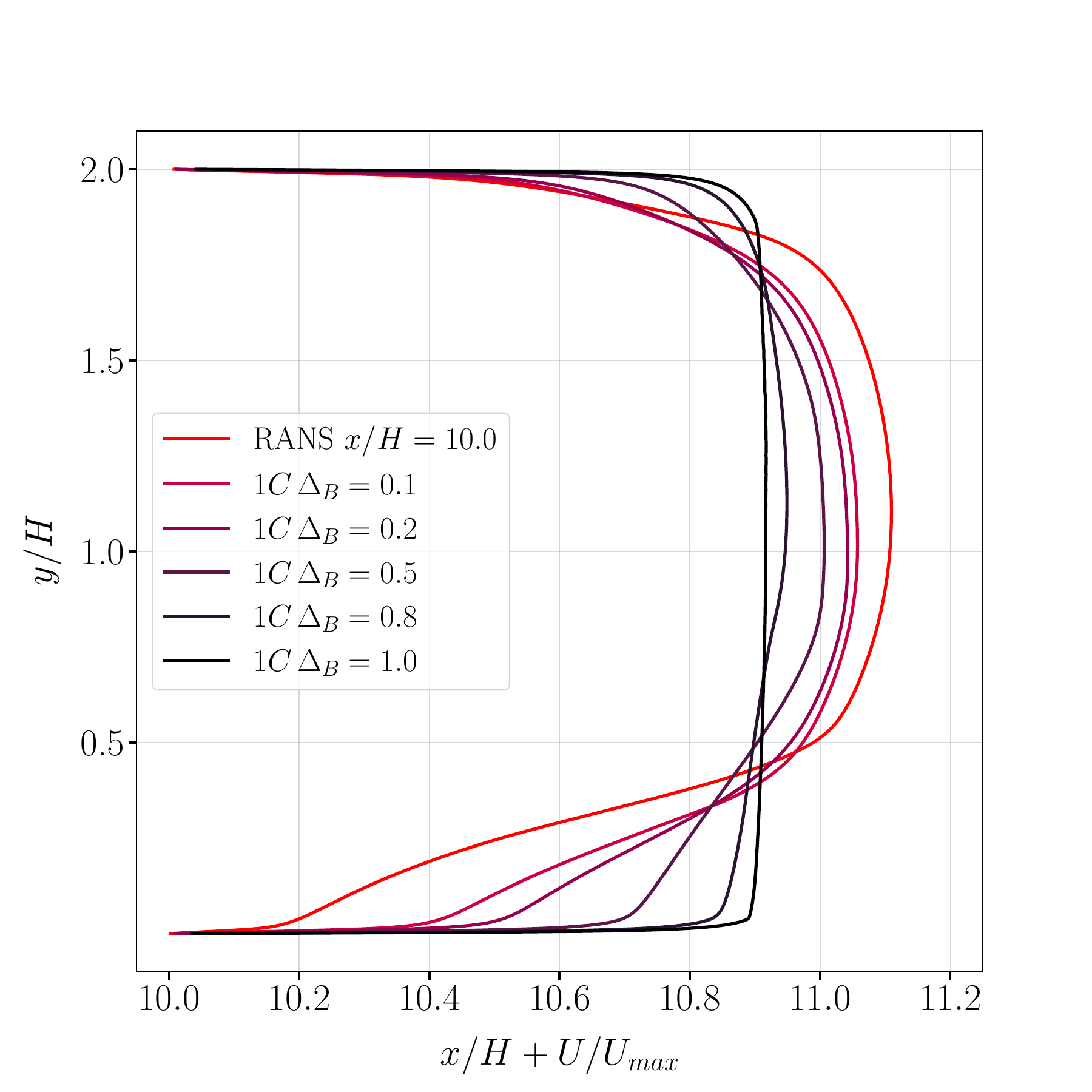}
    };
\node[] at (10.6,4.15) {
     \includegraphics[width=0.495\textwidth, height=8cm, trim=0.4cm  0.2cm 3cm 2cm, clip=True]{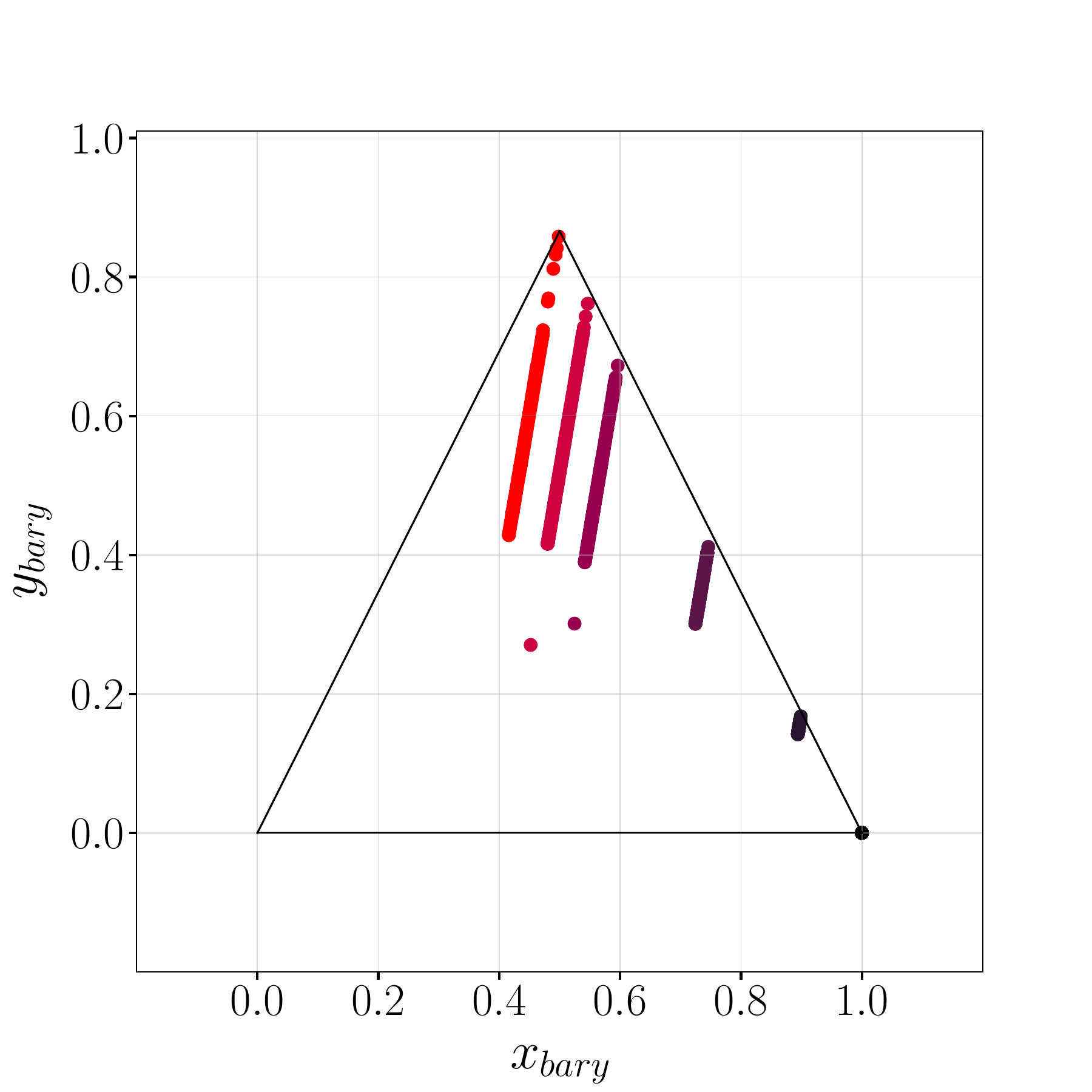}
    };
\node[] at (8.6,2.0) {\small{$\boldsymbol{x}_{2C}$}};
\node[] at (13.5,2.0) {\small{$\boldsymbol{x}_{1C}$}};
\node[] at (11.0,7.3) {\small{$\boldsymbol{x}_{3C}$}};
\draw [fill=white, white] (0.1,0.0) rectangle (0.6,9.0);
\draw [fill=white, white] (7.15,0.0) rectangle (7.6,9.0);

\draw [fill=white, white] (2.0,0.0) rectangle (7.0,0.65);
\draw [fill=white, white] (10.0,0.0) rectangle (15.0,0.65);

\node[] at (3.8,0.4) {\footnotesize{$x/H + U/U_{max}$}};
\node[] at (11.0,0.4) {\footnotesize{$x_{\mathrm{bary}}$}};
\node[rotate=90] at (0.3,4.9) {\footnotesize{$y/H$}};
\node[rotate=90] at (7.4,4.9) {\footnotesize{$y_{\mathrm{bary}}$}};

\end{tikzpicture}
    \caption[Velocity profiles of the $1C$ perturbation and modified barycentric coordinates]{\small{Left: Velocity profiles of the $1C$ perturbation with respect to $\Delta_B$ at $x/H=10.0$, Right: Barycentric coordinates of the $1C$ perturbation with respect to $\Delta_B$ at $x/H=10.0$.}}
    \label{fig:uq_band_delta}
\end{figure}
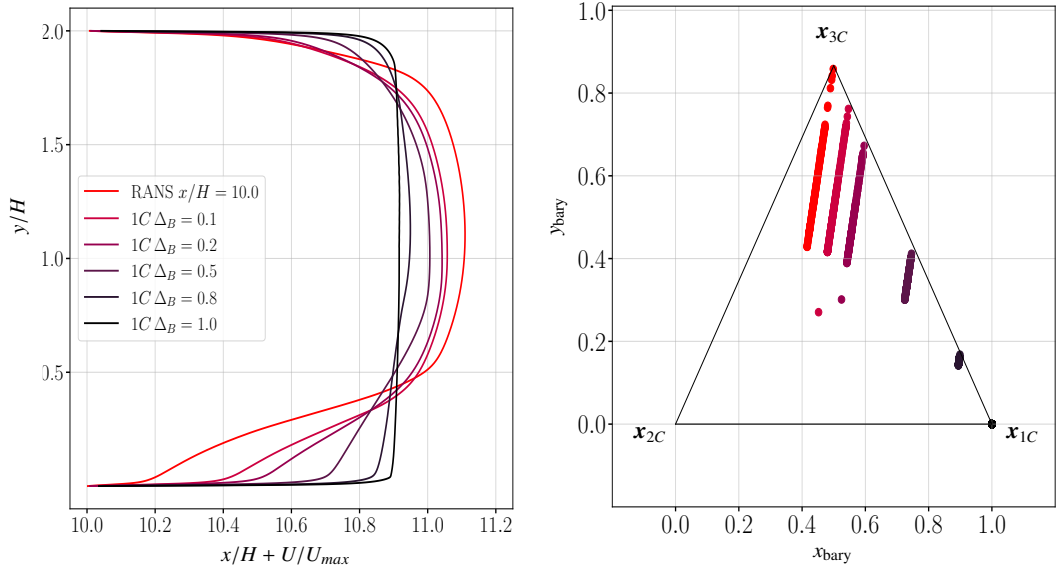

Hence, an increase of the perturbation strength provides a greater (more conservative) uncertainty band. Additionally, an increase of $\Delta_B$ causes a stronger deformation of the velocity profile, thus the uncertainty band may contain regions, which do not enclose the baseline RANS solution at all (see Figure \ref{fig:uq_band_delta}).

\subsection{Data-Driven Approach}
\label{sec:data-driven}
In this chapter, we present the data-driven uncertainty quantification extension introduced by Heyse et al. \cite{heyse2021}. The data-driven method differs from the data-free method by the choice of the (relative) perturbation strength $\Delta_B$. Instead of choosing $\Delta_B$ uniformly distributed in the entire computational domain, machine learning methods should help to predict locally varying perturbation strengths. We use random forests included in the python library \textit{scikit-learn} \cite{pedregosa2011} as the underlying machine learning method in this work.
The selection and compuation of input feature for the machine learning model are similar to our previous work (see Matha et al. \cite{matha2022}).

\subsubsection{Definition of the local perturbation strength}

The fundamental idea is to select the absolute distance in terms of barycentric coordinates of scale-resolving simulations to the ones based on RANS simulations as an indicator for the appropriate amount of perturbation strength
\begin{equation}
    p = ||\boldsymbol{x}_{RANS} - \boldsymbol{x}_{data}||.
\end{equation}
\begin{figure}
\centering
\begin{tikzpicture}
\draw [fill=white, white] (-3,-2) rectangle (3,2.9);
\draw [-](-2.5,-1.6) -- (2.5,-1.6) node[] {};
\draw [-](-2.5,-1.6) -- (0,2.73) node[] {};
\draw [-]( 2.5,-1.6) -- (0,2.73) node[] {};
\draw [gray, dashed]( -0.3, 0.9) -- (0,2.73) node[] {};
\draw [gray, dashed]( -0.3, 0.9) -- (-2.5,-1.6) node[] {};
\draw [gray, dashed]( -0.3, 0.9) -- (2.5,-1.6) node[] {};
\draw [dashed](-0.3, 0.9) circle (1.1cm);
\draw [red]( -0.3, 0.9) -- (-0.3, -0.2) node[] {};
\filldraw[black] (-0.26,0.9) circle (2pt) node[anchor=west]{$\boldsymbol{x}$};
\filldraw[black] (-2.3, -1.6) circle (0.01pt) node[anchor=north east]{$\boldsymbol{x}_{2C}$};
\filldraw[black] (2.3, -1.6) circle (0.01pt) node[anchor=north west]{$\boldsymbol{x}_{1C}$};
\filldraw[black] (0, 2.73) circle (0.01pt) node[anchor=south]{$\boldsymbol{x}_{3C}$};
\filldraw[black] (-0.3, -0.2) circle (2pt) node[anchor=north]{$\boldsymbol{x}_{data}$};
\filldraw[black] (-1.03, 0.06) circle (1pt) node[]{};
\filldraw[black] (0.51, 0.16) circle (1pt) node[]{};
\filldraw[black] (-0.12, 1.98) circle (1pt) node[]{};
\draw [-]( -1.03, 0.06) -- (-0.3, 0.9) node[] {};
\draw [-]( 0.51, 0.16) -- (-0.3, 0.9) node[] {};
\draw [-]( -0.12, 1.98) -- (-0.3, 0.9) node[] {};
\node[red] at (-0.1, 0.35) {$p$};
\end{tikzpicture}
    \caption[Perturbation of the barycentric coordinates using the data-driven method]{\small{Perturbation of the barycentric coordinates using the data-driven method.}}
    \label{fig:barycentric_data}
\end{figure}
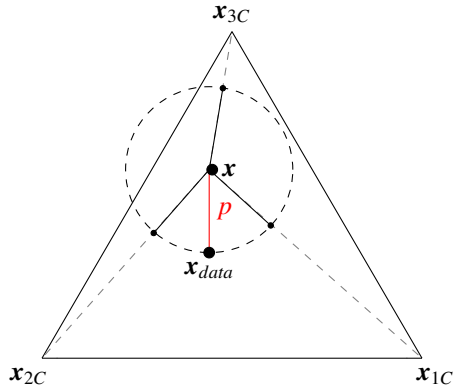

Furthermore, similar to the data-free approach, the three vertices act as target points for the perturbations. Hence, every RANS determined state $\boldsymbol{x}$ is shifted by the radius $p$ around $\boldsymbol{x}$ towards on of the limiting corners of the triangle (see Figure \ref{fig:barycentric_data}). 

One of the main differences compared to the data-free method is the type of perturbation. For the data-free approach, each point is shifted relatively to the vertex. For the data-driven application, the perturbation strength is given as an absolute value. Convergence issues can also arise when using the data-driven approach (mostly for the $3C$ perturbation), which requires the application of the moderation factor $f$, when reconstructing the perturbed Reynolds stress tensor $\tau^*$ (see Equation \eqref{gl:perturbatedReStress}).

While the data-driven idea is motivated by avoiding too conservative uncertainty estimates, it has to be clarified if the choice of a local perturbation strength provides an improvement for the eigenspace perturbation framework from a theoretical point of view.  

\subsubsection[Estimating the Perturbation Magnitude]{Estimating the perturbation magnitude}
A classical methodology for training a machine learning model is to divide data sets into training, validation and test data sets. The model is trained with the data of the training data set. The hyperparameters for the learning algorithm are determined using the validation data set. For this purpose, the models are trained with different hyperparameters. The hyperparameters that provide the best accuracy of the model on the validation data set are chosen for the final model. Finally, the model is evaluated on the test data set to examine the model's performance on data, that has not been used for training or validation.

Firstly, we train the random forests on four of the five extant Reynolds number channel flow cases based on DNS simulations of Lee and Moser \cite{lee2015} and test it on the remaining fifth data set. We do not perform a validation with a subset of the training data, but choose the best possible hyperparameters evaluated on the fifth data set. This is in contrast to common procedures, since the hyperparameters are adapted to the test data set hereby. This is referred to as k-fold cross validation. However, we will see that even choosing the best possible hyperparameters does not lead to satisfactory results.

Secondly, we analyze how well the model performs on a new geometry. For this purpose, several data sets are selected for training (DNS of turbulent channel flow at $Re_\tau \in \{180, 550, 1000,\\
2000, 5200\}$ \cite{lee2015},  DNS at $Re_H \in \{2800, 5600\}$ and LES at $Re_H = 10595$ of periodic hill flow \cite{breuer2009}, DNS of wavy wall flow at $Re_H = 6850$ \cite{rossi2006}, while the resulting random forest is tested on the converging-diverging channel \cite{laval2011}. Hereby, we choose the best possible hyperparameters based on the data of the converging-diverging channel.

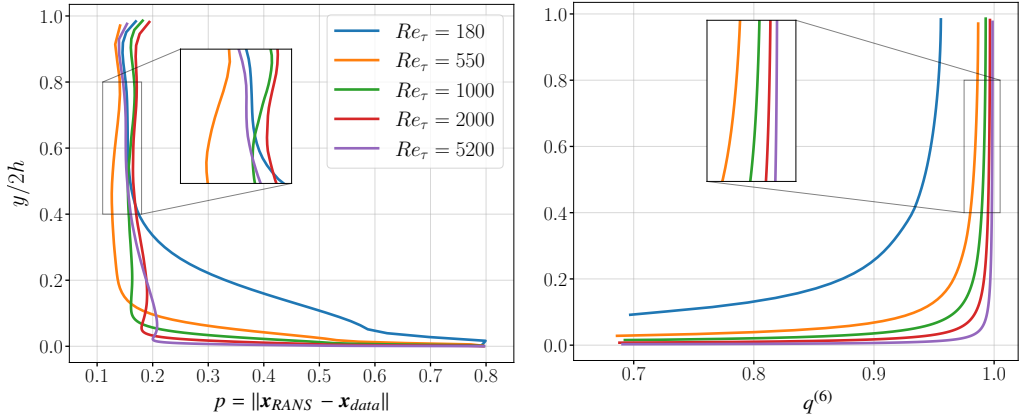
\begin{figure}[!htb]
\centering
\begin{tikzpicture}
\node[anchor=south west] (image) at (0,0) {
     \includegraphics[width=0.495\textwidth,trim=0.5cm 0.35cm 3cm 0.42cm, clip=True] {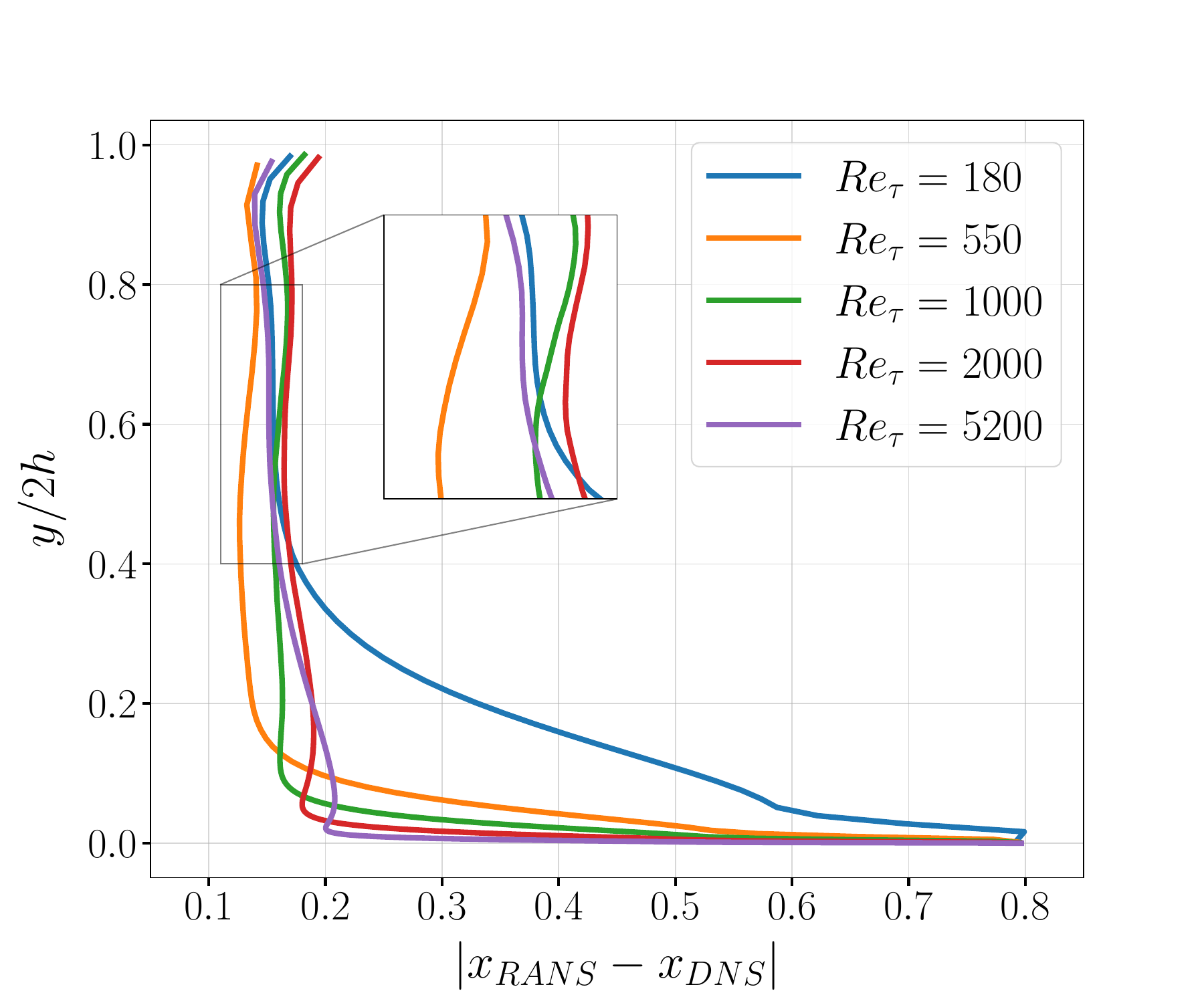}
    };
\node[] at (10.1,3.2) {
     \includegraphics[width=0.495\textwidth,trim=0.5cm 0.35cm 3cm 0.42cm, clip=True]{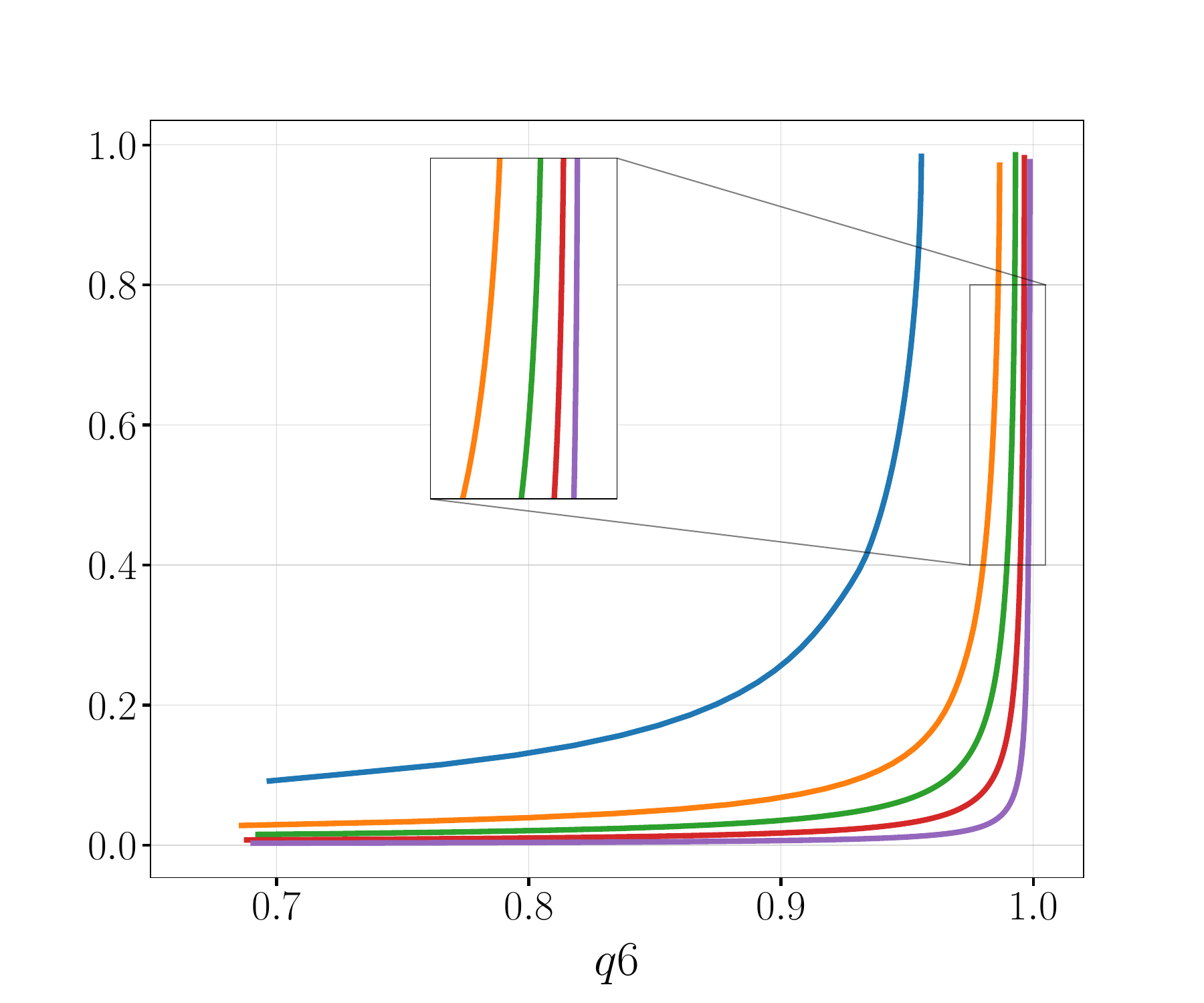}
    };
\draw [fill=white,white] (2,0.1) rectangle (6.5,0.5);
\node[] at (4,0.3) {\footnotesize $p=||\boldsymbol{x}_{RANS}-\boldsymbol{x}_{data}||$};
\draw [fill=white,white] (8.4,0.1) rectangle (18.5,0.5);
\node[] at (10.8,0.3) {\footnotesize{$q^{\left(6\right)}$}};
\end{tikzpicture}
    \caption[Radius of perturbation and feature $q^{\left(6\right)}$ of the turbulent channel flows]{\small{Left: Radius of perturbation for different Reynolds numbers of the turbulent channel flow, Right: Distribution of the feature $q^{\left(6\right)} = P_k / \left(|P_k| + |k \omega|\right)$. }}
    \label{fig:kanal_data}
\end{figure}

\paragraph[Channel Flow Example]{Channel flow example}
In order to analyze the generalization property of machine learning models trained with flow data of the same geometry, but different Reynolds numbers, we select simple turbulent channel flow. The distributions of the target quantity for the different data sets are shown on the left in Figure \ref{fig:kanal_data}. The general shape of the perturbation radius distribution is similar for each Reynolds number.

We observe the non-linear dependence of the target quantity as a function of the Reynolds number. The perturbation radius of the Reynolds number $Re_\tau=550$ is smallest in the center of the channel, whereas the perturbation radius of the Reynolds number $Re_\tau=2000$ is largest in this region. Unfortunately, this does not apply to the absolute value of certain features. For example, feature $q^{\left(6\right)}$, related to the turbulent production of the turbulent kinetic energy equation, is shown on the right in Figure \ref{fig:kanal_data}. The larger the Reynolds number, the larger the corresponding feature. Nevertheless, it can also be seen that there is a correlation between feature $q^{\left(6\right)}$ and the perturbation radius. The smaller the feature $q^{\left(6\right)}$, the larger the perturbation strength and vice versa. However, the linearity of the absolute value of the feature and the non-linearity of the target quantity lead to impossible correct machine learning local prediction. In particular, this is presented in the following study.

The channel flow data based on $Re_\tau=180, 550, 2000, 5200$ is used as a training data set and the model is evaluated on $Re_\tau=1000$. A hyperparameter grid search is performed for each model ($n_{estimators} \in \{2, 5, 10, 20, 30, 40, 50, 75, 100\}$, $\max_{depth} \in \{2 , 5, 7, 10, 15\}$ and $\max_{features} \in \{2, 4, 6, 8\}$). Moreover, we include the data of $Re_\tau = 1000$ in the training data set and apply the model based on this for $Re_\tau=550$ (test data set). Rather than discussing the results of the hyperparameter search, we present the errors based on the best hyperparameters in Table \ref{tab:kanal}.

\begin{table}[t]
    \centering
    \footnotesize
    \begin{tabular}{ |c|c|c|c|c|c| } 
\hline
Case & Training & Test & RMSE Training & RMSE Test & Hyperparameter \\
\hline
\multirow{4}{3em}{\centering 1} & \multirow{1}{8em}{\centering $Re_\tau = \{180$} & \multirow{4}{6em}{\centering$Re_\tau=1000$} & \multirow{4}{3em}{\centering0.012}& \multirow{4}{3em}{\centering 0.015} & \multirow{1}{8em}{\centering$n_{estimators}=30$} \\ 
&550  & & & &$\max_{depth}=5$  \\ 
&2000  & & & &$\max_{features}=6$ \\ 
&$5200\}$  & & & &  \\ 
\hline
\multirow{4}{3em}{\centering 2} & \multirow{1}{8em}{\centering $Re_\tau=\{180$} & \multirow{4}{5em}{\centering $Re_\tau=550$} & \multirow{4}{3em}{\centering0.012}& \multirow{4}{3em}{\centering 0.030} & \multirow{1}{8em}{\centering$n_{estimators}=20$} \\ 
&1000  & & & &$\max_{depth}=5$  \\ 
&2000  & & & &$\max_{features}=4$ \\ 
&$5200\}$  & & & &  \\ 
\hline
\end{tabular}
    \caption[Results of the machine learning models for the channel flow test cases]{\small{Results of the machine learning models for the channel flow use cases.}}
    \label{tab:kanal}
\end{table}

\begin{figure}[!htb]
\centering

\begin{tikzpicture}
\node[anchor=south west] (image) at (0,0) {
     \includegraphics[width=0.495\textwidth,trim=0.5cm 0.2cm 3cm 0.42cm, clip=True] {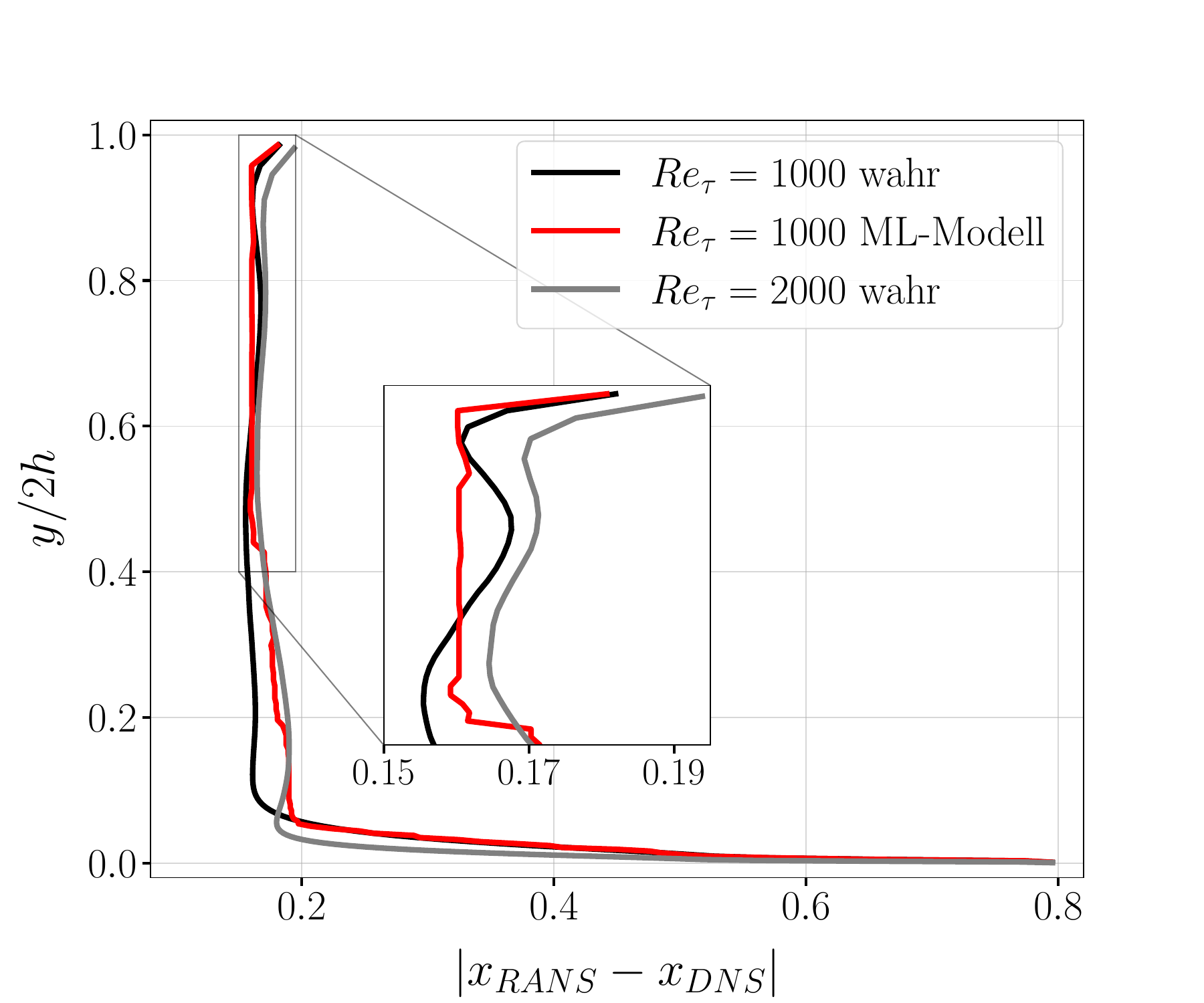}
    };
\node[] at (10.1,3.2) {
     \includegraphics[width=0.495\textwidth,trim=0.5cm 0.2cm 3cm 0.42cm, clip=True]{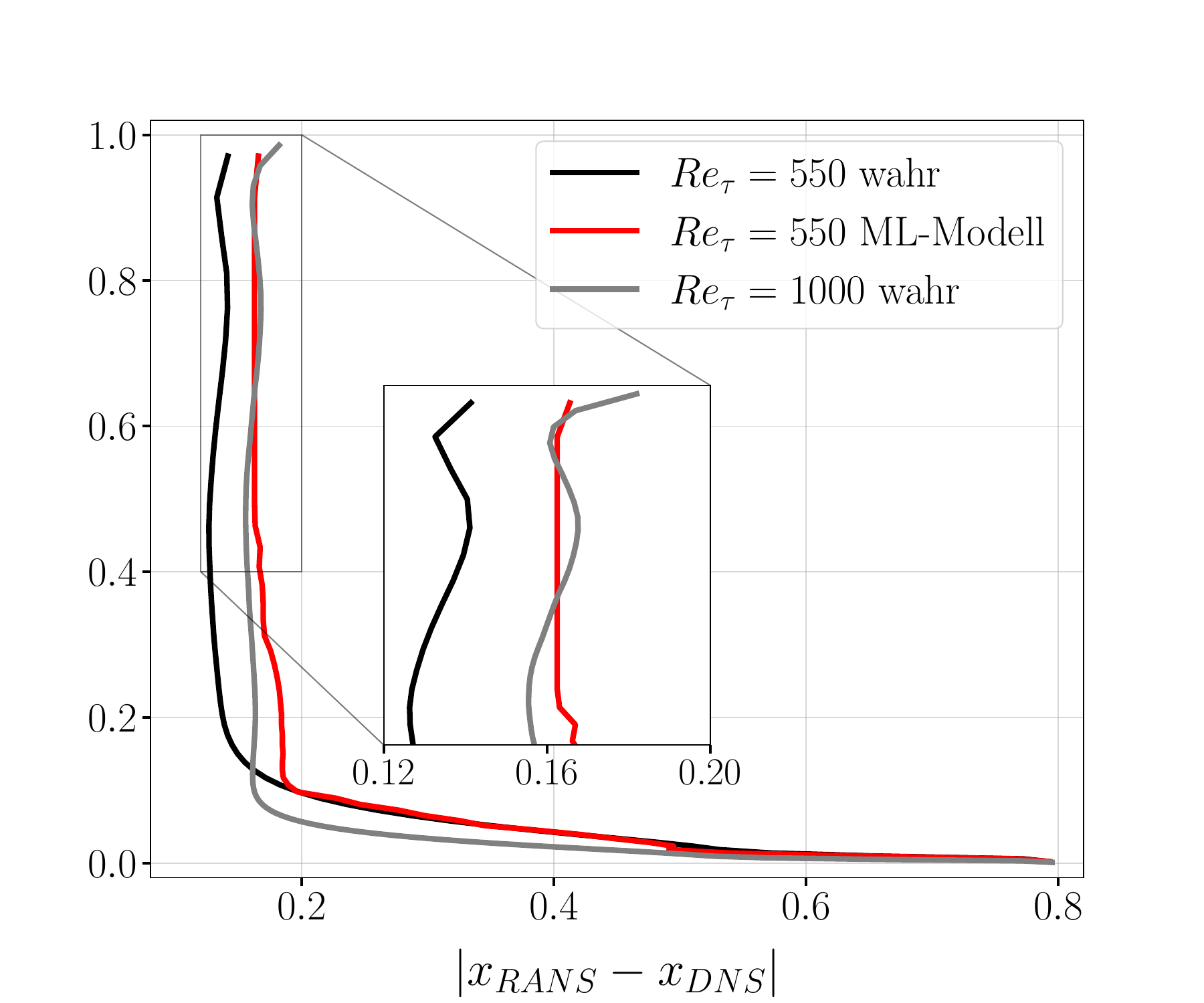}
    };
\draw [fill=white,white] (2,0.1) rectangle (6.5,0.5);
\node[] at (4.2,0.3) {\footnotesize{$p=||\boldsymbol{x}_{RANS}-\boldsymbol{x}_{data}||$}};
\draw [fill=white,white] (6,0.1) rectangle (13.5,0.5);
\node[] at (10.8,0.3) {\footnotesize{$p=||\boldsymbol{x}_{RANS}-\boldsymbol{x}_{data}||$}};
\draw [fill=white,white] (5.4,5) rectangle (6.3,5.4);
\node[] at (5.7,5.28) {\footnotesize true};
\draw [fill=white,white] (5.4,4.4) rectangle (6.3,4.8);
\node[] at (5.7,4.55) {\footnotesize true};
\draw [fill=white,white] (12,5) rectangle (13,5.4);
\node[] at (12.3,5.28) {\footnotesize true};
\draw [fill=white,white] (12.1,4.4) rectangle (13,4.8);
\node[] at (12.4,4.55) {\footnotesize true};
\end{tikzpicture}

    \caption[Machine learning models predictions for the
     channel flow test cases]{\small{Prediction of the machine learning models when data for one Reynolds number is not included in the training data set. Left: $Re_\tau=1000$ not included in the training data set, Right: $Re_\tau=550$ not included in the training data set.}}
    \label{fig:kanal_ml}
\end{figure}
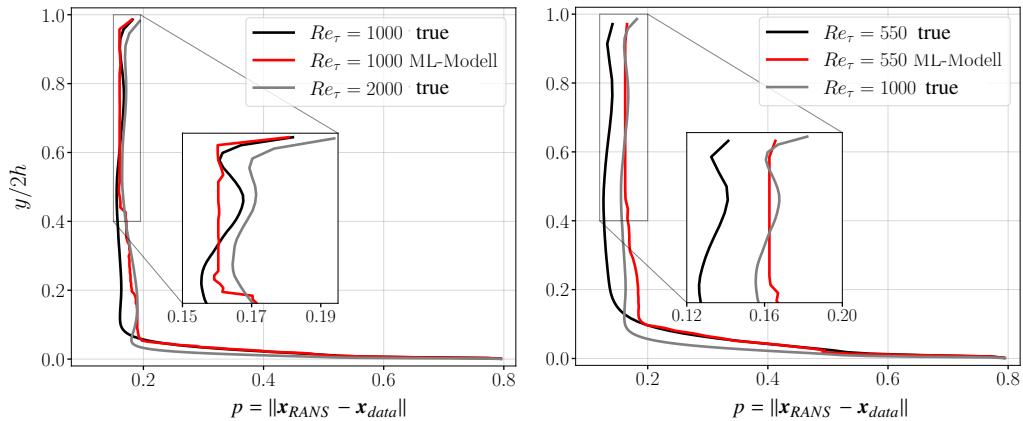

By verifying the predicted distribution of the perturbation radius in Figure \ref{fig:kanal_ml}, we notice that the random forest is able to predict the general shape of the distribution well. However, it can also be seen that for $0.1 < y/2h < 0.4$ the prediction is closer to the perturbation strength of $Re_\tau=2000$ when actually trying to predict $Re_\tau=1000$. 

For testing on $Re_\tau=550$, the shape of the curve is adequate approximated as well. But for $y/2h > 0.1$ the predicted distribution is close to $Re_\tau=1000$. This is due to the non-linearity of the target quantity, shown above. Since the quantity of almost all features for $Re_\tau=550$ are between the features of $Re_\tau=180$ and $Re_\tau=1000$, the prediction for the perturbation radius is also in between.

Thus, in general it is possible to learn and predict the perturbation radius for the channel flow based on different Reynolds numbers. The random forest benefits from strong similarity of the target values. Nevertheless, the predicted perturbation radius is often very similar to the perturbation radius of a channel flow from the training data set. This is not due to overfitting, as the hyperparameters are chosen to reduce the errors with respect to test data sets. 

\paragraph[Converging-diverging channel]{Converging-diverging channel example} In this study, we investigate the generalization property of the models with respect to a new geometry. Data of the turbulent channel, the periodic hill and the wavy wall serve as training data. We evaluate the trained model in predicting the perturbation radius for the converging-diverging channel. 

\begin{figure}[!htb]
\centering
\begin{tikzpicture}
\node[anchor=south west] (image) at (0,0) {
    \includegraphics[width=\textwidth]{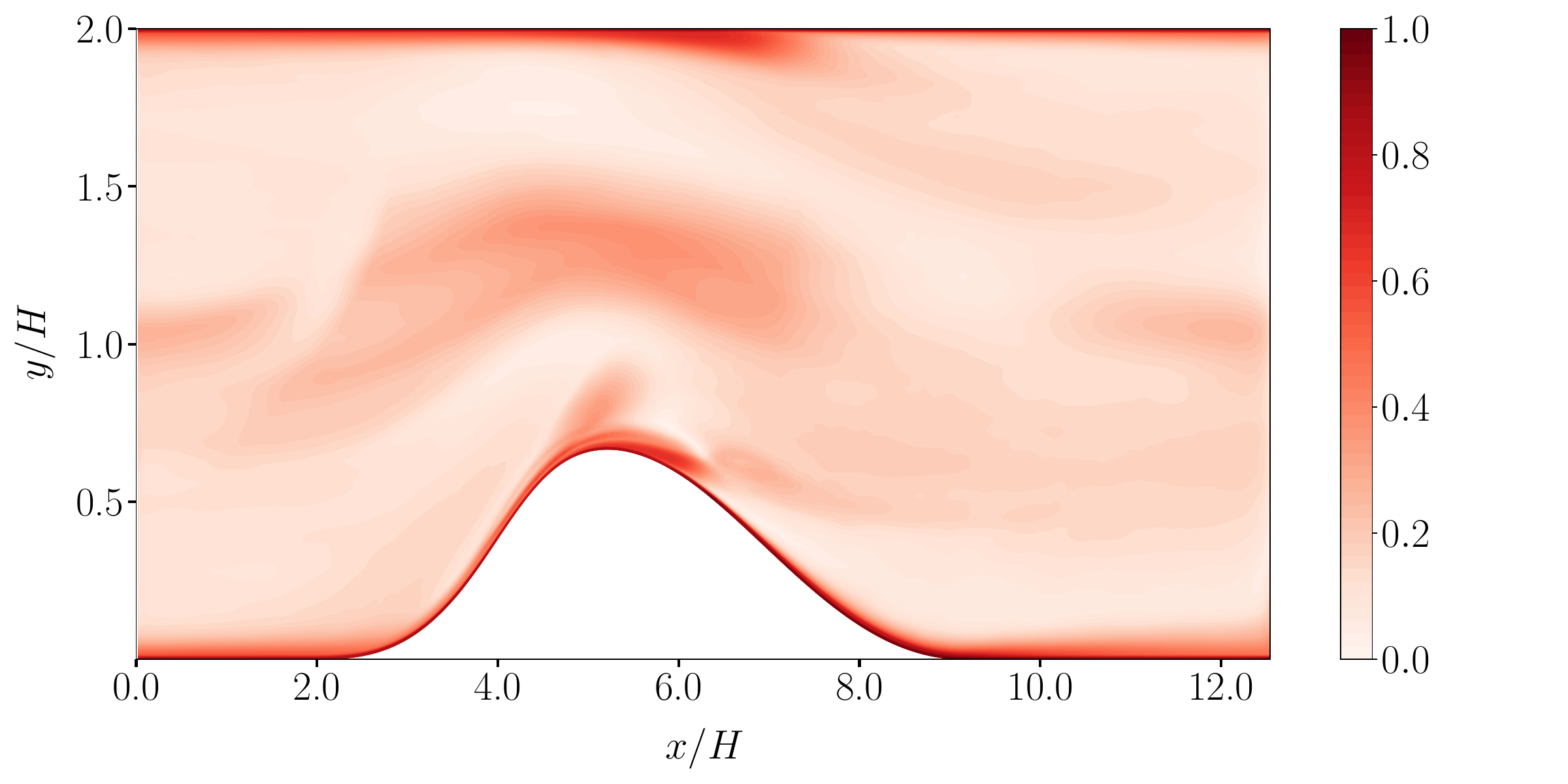}
    };
\draw[rounded corners=0.8mm] (1.32, 3.3) rectangle (2.9, 4.6);
\node[] at (2.1, 4.9) {$1$};
\draw[rounded corners=0.8mm] (9, 3.5) rectangle (10.95, 4.8);
\node[] at (10, 5.1) {$3$};
\draw[rounded corners=0.8mm] (3.1, 3.5) rectangle (8, 5.5);
\node[] at (5.55, 5.8) {$2$};
\node[] at (12.8, 3.9) {$p$};
\end{tikzpicture}
    \caption[Distribution of the perturbation radius of the converging-diverging channel]{\textup{\small{Distribution of the perturbation radius for the converging-diverging channel.}}}
    \label{img:perRad_conv_div}
\end{figure}

The correct perturbation radius $p$ of the converging-diverging channel, which can be determined as DNS data of Laval and Marquillie in \cite{laval2011} exists, is shown in Figure \ref{img:perRad_conv_div}. Similar to the channel flow, $p$ increases near the wall. The perturbation radius in region 2 in Figure \ref{img:perRad_conv_div} is larger than the perturbation radius of the surrounding points. Areas, featuring a greater perturbation radius can also be identified at the inlet and outlet at $y/H\approx1.0$ (areas 1 and 3). 

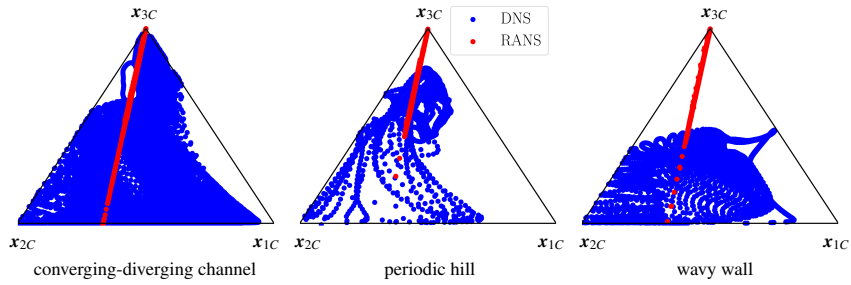
\begin{figure}[htb]
\centering
\begin{tikzpicture}
\node[anchor=south west] (image) at (0,0) {
     \includegraphics[width=\textwidth]{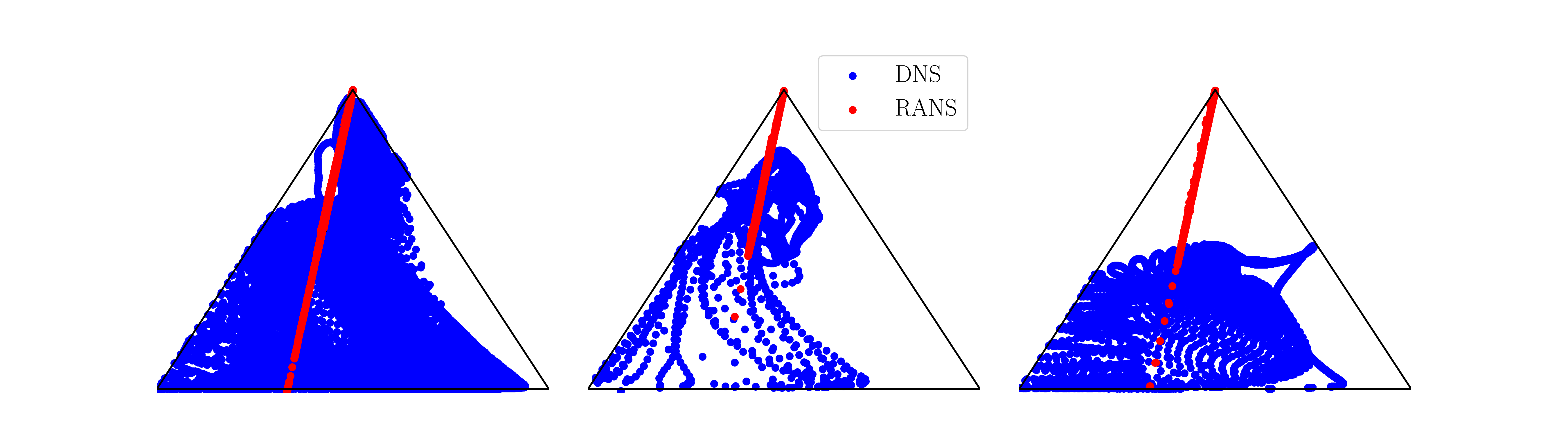}
    };
\node[] at (1.55, 0.25) {\scriptsize{$\boldsymbol{x}_{2C}$}};
\node[] at (4.75, 0.25) {\scriptsize{$\boldsymbol{x}_{1C}$}};
\node[] at (5.3, 0.25) {\scriptsize{$\boldsymbol{x}_{2C}$}};
\node[] at (8.5, 0.25) {\scriptsize{$\boldsymbol{x}_{1C}$}};
\node[] at (9.05, 0.25) {\scriptsize{$\boldsymbol{x}_{2C}$}};
\node[] at (12.25, 0.25) {\scriptsize{$\boldsymbol{x}_{1C}$}};
\node[] at (3.15, 3.3) {\scriptsize{$\boldsymbol{x}_{3C}$}};
\node[] at (6.9, 3.3) {\scriptsize{$\boldsymbol{x}_{3C}$}};
\node[] at (10.65, 3.3) {\scriptsize{$\boldsymbol{x}_{3C}$}};
\node[] at (3.15, -0.1) {\scriptsize{converging-diverging channel}};
\node[] at (6.9, -0.1) {\scriptsize{periodic hill}};
\node[] at (10.65, -0.1) {\scriptsize{wavy wall}};
\end{tikzpicture}
    \caption[Distribution of barycentric coordinates ]{\small{Distribution of barycentric coordinates of the DNS and RANS calculations.}}
    \label{fig:verteilung_triagnle_2d}
\end{figure}

\begin{figure}[htb]
\begin{center}
\mbox{
\begin{tikzpicture}
\node[anchor=south west] (image) at (0,0) {
     {\includegraphics[scale=0.152, trim=4.65cm 2.5cm 0.1cm 0.1cm, clip=True]{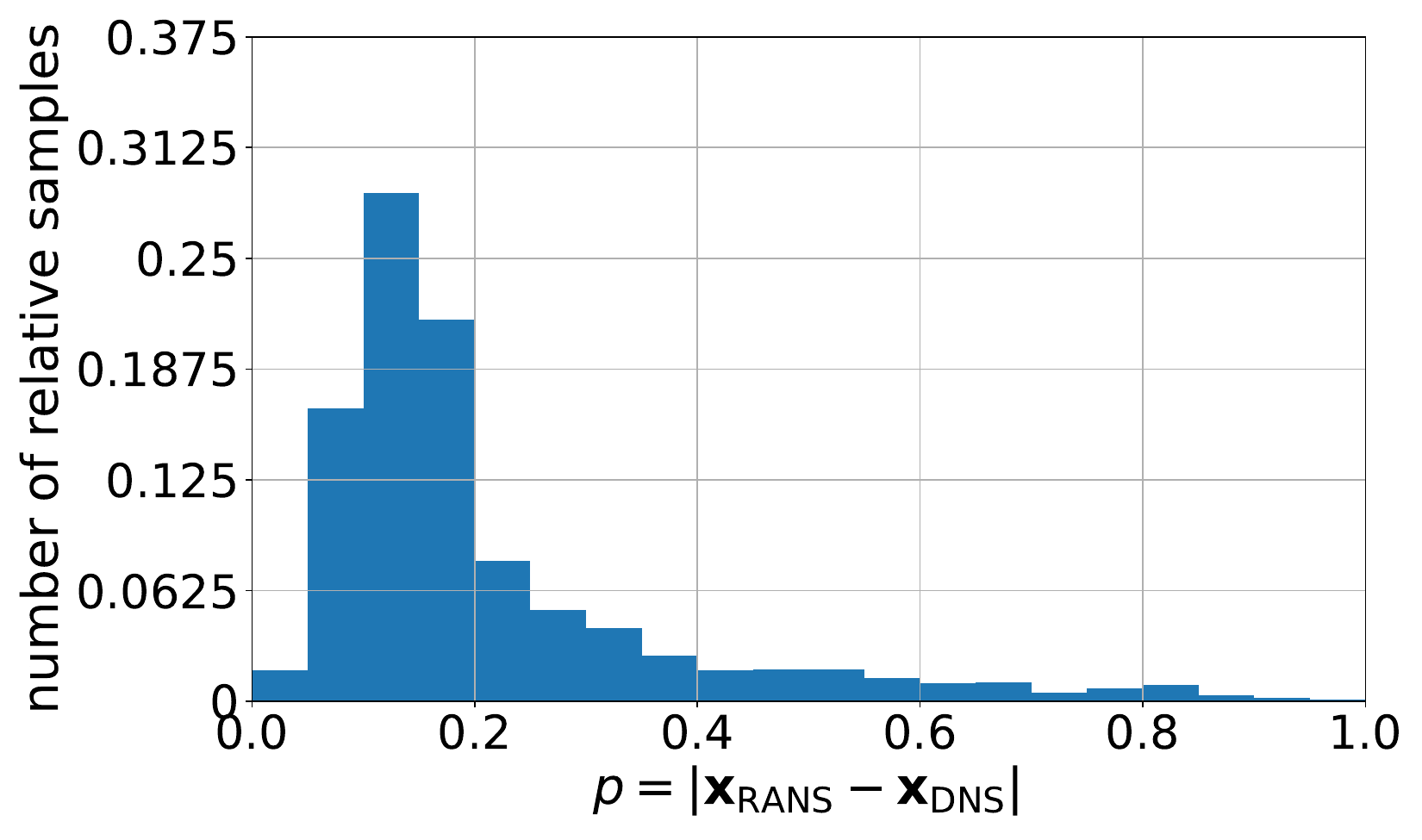}\label{histoconvDivbary}}
    };
    \node[] at (-0.25, 0.51) {\tiny{$0.0625$}};
    \node[] at (-0.25, 0.84) {\tiny{$0.1250$}};
    \node[] at (-0.25, 0.84+0.32) {\tiny{$0.1875$}};
     \node[] at (-0.25, 0.84+0.32+0.32) {\tiny{$0.2500$}};
      \node[] at (-0.25, 0.84+0.32+0.32+0.32) {\tiny{$0.3125$}};
       \node[] at (-0.25, 0.84+0.32+0.32+0.32+0.32) {\tiny{$0.3750$}};
       \node[] at (0.18, 0.0) {\tiny{$0.0$}};
       \node[] at (0.83, 0.0) {\tiny{$0.2$}};
       \node[] at (0.83+0.65, 0.0) {\tiny{$0.4$}};
       \node[] at (0.83+0.65+0.65, 0.0) {\tiny{$0.6$}};
       \node[] at (0.83+0.65+0.65+0.65, 0.0) {\tiny{$0.8$}};
       \node[] at (0.83+0.65+0.65+0.65+0.65, 0.0) {\tiny{$1.0$}};
        \node[] at (1.8, -0.2) {\tiny{$p$}};
        \node[rotate=90] at (-0.7, 1.2) {\tiny{Relative frequency}};
        \node[] at (1.8, -0.5) {\scriptsize{converging-diverging channel}};
\end{tikzpicture}
}
\mbox{
\begin{tikzpicture}
\node[anchor=south west] (image) at (0,0) {
     {\includegraphics[scale=0.15
, trim=0.1cm 2.5cm 0cm 0cm, clip=true]{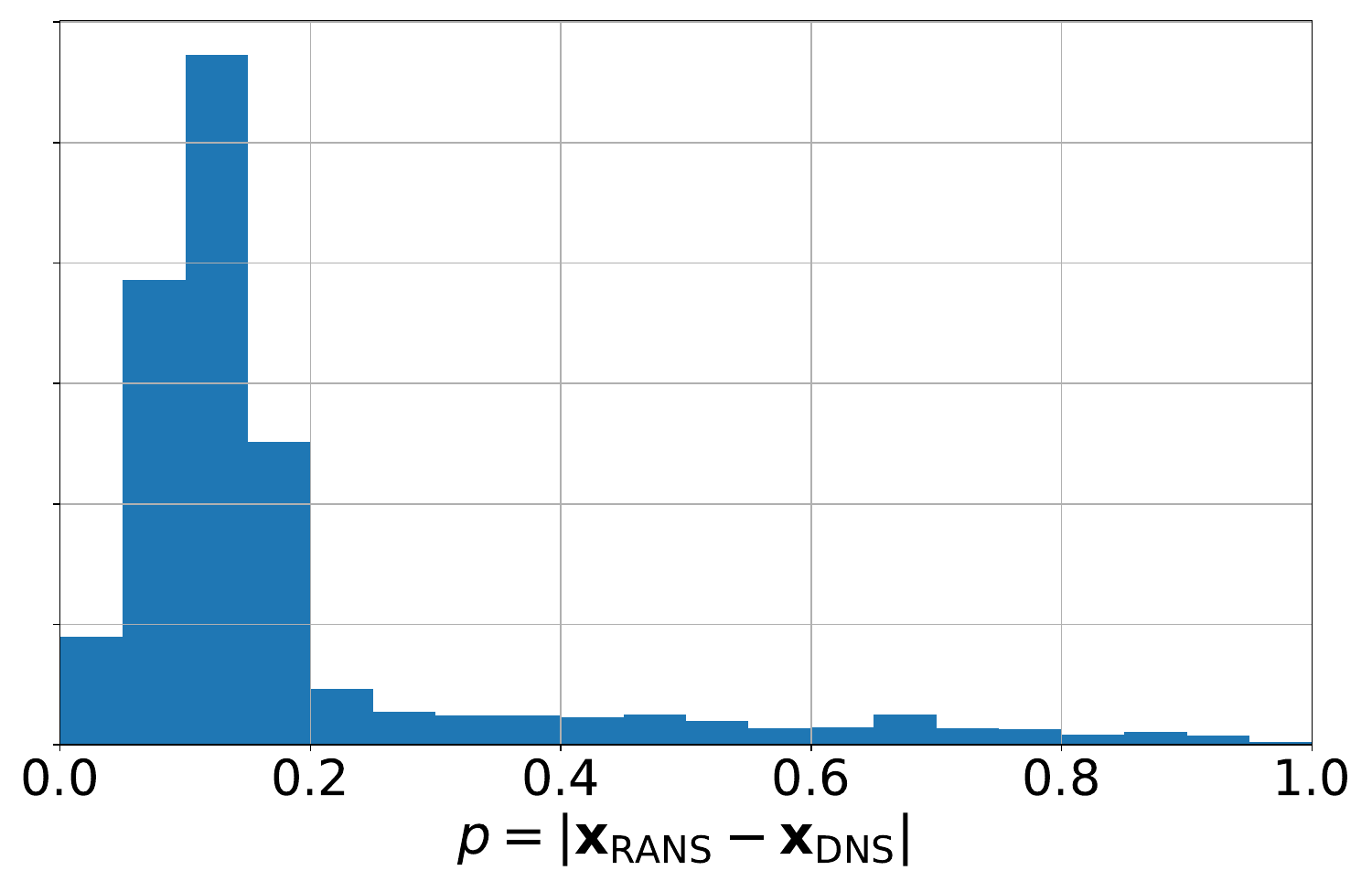}\label{histohillbary}}
    };
    \node[] at (0.28, 0.0) {\tiny{$0.0$}};
       \node[] at (0.93, 0.0) {\tiny{$0.2$}};
       \node[] at (0.98+0.65, 0.0) {\tiny{$0.4$}};
       \node[] at (1.02+0.65+0.65, 0.0) {\tiny{$0.6$}};
       \node[] at (1.04+0.65+0.65+0.65, 0.0) {\tiny{$0.8$}};
       \node[] at (1.09+0.65+0.65+0.65+0.65, 0.0) {\tiny{$1.0$}};
        \node[] at (2.0, -0.2) {\tiny{$p$}};
        \node[] at (2, -0.5) {\scriptsize{periodic hill}};
\end{tikzpicture}
}
\mbox{
\begin{tikzpicture}
\node[anchor=south west] (image) at (0,0) {
     {\includegraphics[scale=0.15, trim=0.1cm 2.5cm 0cm 0cm, clip=true]{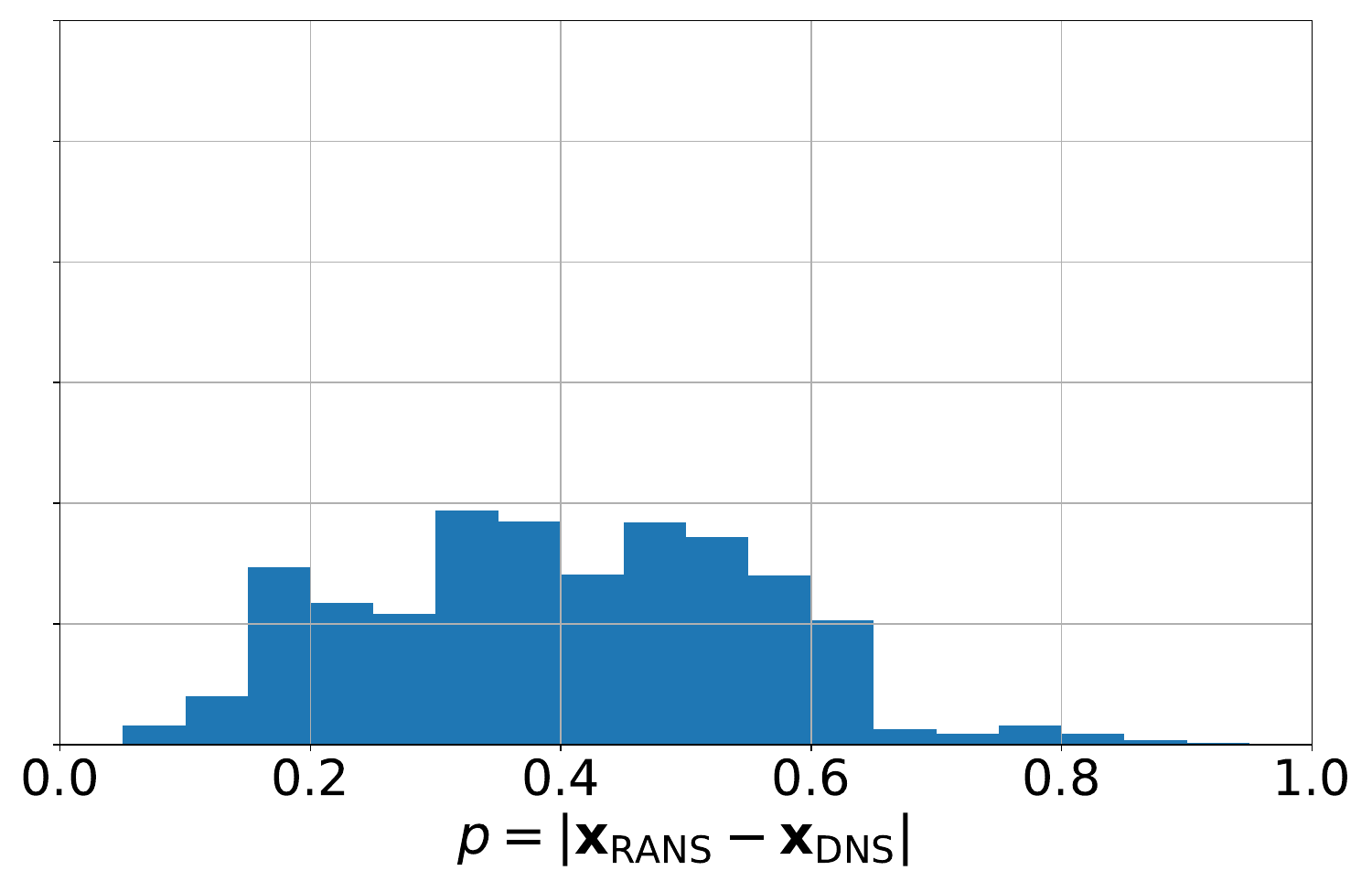}\label{histohillWavy}}
    };
    \node[] at (0.28, 0.0) {\tiny{$0.0$}};
       \node[] at (0.93, 0.0) {\tiny{$0.2$}};
       \node[] at (0.98+0.65, 0.0) {\tiny{$0.4$}};
       \node[] at (1.02+0.65+0.65, 0.0) {\tiny{$0.6$}};
       \node[] at (1.04+0.65+0.65+0.65, 0.0) {\tiny{$0.8$}};
       \node[] at (1.09+0.65+0.65+0.65+0.65, 0.0) {\tiny{$1.0$}};
        \node[] at (2.0, -0.2) {\tiny{$p$}};
        \node[] at (2, -0.5) {\scriptsize{wavy wall}};
\end{tikzpicture}
}
\caption[Relative frequency of the perturbation radius for the 2D flow data sets]{\small{Relative frequency of the perturbation radius of the data points from Figure \ref{fig:verteilung_triagnle_2d}.}}
\label{fig:verteilung_p}
\end{center}
\end{figure}
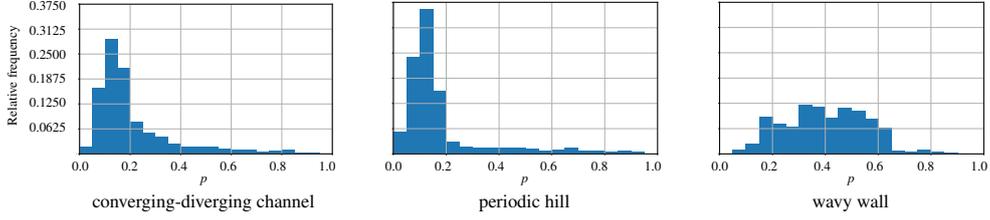

The relative frequency of the barycentric coordinates of flow data sets is outlined in Figure \ref{fig:verteilung_triagnle_2d}. The barycentric coordinate distribution of the wavy wall differs from the barycentric coordinate distribution of the other two data sets. The barycentric coordinates of the wavy wall are not located in the upper part of the triangle. Thus, the perturbation radius of the wavy wall has to be greater compared to the perturbation radius of the other two data sets (see Figure \ref{fig:verteilung_p}). Consequently, is becomes obvious that a model trained on the wavy wall data predicts greater perturbation radii for the converging diverging channel.

In order to shed a light on this, different training data sets are used to train a random forest and evaluate the prediction on the converging-diverging channel (see Table \ref{tab:ml_conv_div}). 
The hyperparameters are chosen such that the final model provides the smallest errors predicting the target quantity for the converging-diverging channel. As already described above, this is not a common procedure in practice, since the model is directly adapted to the test data set. However, we are interested in the best performing models based on the available data sets. The errors of the best models and the selected hyperparameters are shown in Table \ref{tab:ml_conv_div}. The grid search was performed with the parameters $n_{estimators} \in \{5, 10, 20, 30, 40, 50, 70, 100\}$, $\max_{depth} \in \{2, 5 , 7, 10, 15\}$ and $\max_{features} \in \{2, 4, 6, 8\}$.

The prior conjecture is reflected in the errors of the model predictions. Adding the wavy wall data for training purposes worsens the model prediction. When the model is only trained with the data of the channel flow and the periodic hill, the errors on the test data set are lower accordingly.  

\begin{figure}[!t]
\centering
\includegraphics[width=\textwidth]{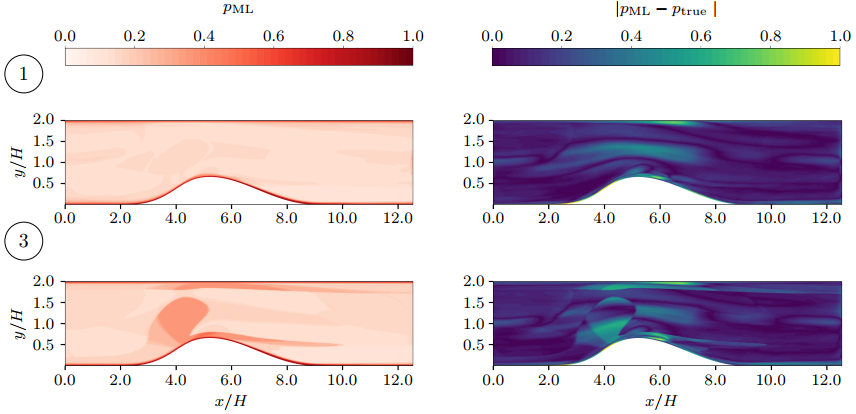}
    \caption[Prediction of the ML models for the converging-diverging channel]{\small{Left: Prediction of the machine learning models from $p$ for cases 1 and 3 from Table \ref{tab:ml_conv_div}, Right: Absolute errors of the predictions by $p$.}}
    \label{fig:ml_fall_2_mit_fehler}
\end{figure}

\begin{figure}[!htb]
\centering
\includegraphics[width=\textwidth]{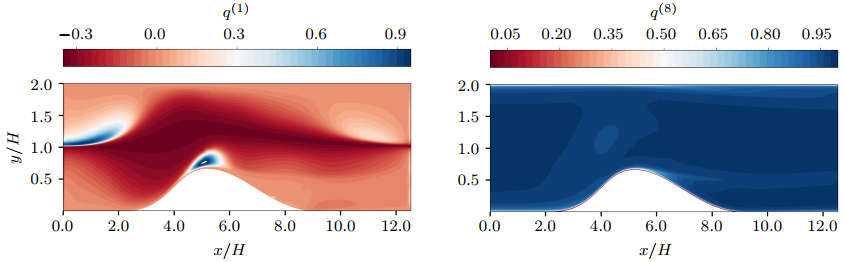}
    \caption[Distribution of Feature $q^{\left(1\right)}$ and $q^{\left(8\right)}$ of converging-diverging channel]{\small{Distribution of Feature $q^{\left( 1\right)} = \left(0.5||\Omega||^2-0.5||S||^2\right)/ \left[\left(0.5||\Omega||^2-0.5||S||^2\right) + ||S||^2 \right]$ and $q^{\left(8\right)} = \mu_t / \left(|\mu_t| + |\mu|\right)$ in the converging-diverging channel.}}
    \label{fig:feature}
\end{figure}

\begin{table}[htb]
    \centering
    \footnotesize
    \begin{tabular}{ |c|c|c|c|c| } 
\hline
Fall & Training  & RMSE Training & RMSE Test & Hyperparameter \\
\hline
\multirow{3}{3em}{\centering 1} & \multirow{1}{7em}{\centering channel flow}  & \multirow{3}{3em}{\centering0.025}& \multirow{3}{3em}{\centering 0.076} & \multirow{1}{8em}{\centering$n_{estimators}=70$} \\ 
& periodic hill  & & &{\centering$\max_{depth}=7$}  \\ 
& & & &{\centering$\max_{features}=6$} \\
\hline
\multirow{3}{3em}{\centering 2} & \multirow{1}{7em}{\centering channel flow} & \multirow{3}{3em}{\centering0.038}& \multirow{3}{3em}{\centering 0.083} & \multirow{1}{8em}{\centering $n_{estimators}=20$} \\ 
& periodic hill  & & &{\centering $\max_{depth}=10$}  \\ 
&wavy wall (entirely)  & & &{\centering $\max_{features}=6$} \\ 
\hline
\multirow{3}{3em}{\centering 3} & \multirow{1}{7em}{\centering channel flow}  & \multirow{3}{3em}{\centering0.034}& \multirow{3}{3em}{\centering 0.082} & \multirow{1}{8em}{\centering $n_{estimators}=30$} \\ 
& periodic hill  & & &{\centering $\max_{depth}=10$}  \\ 
&wavy wall (every 2\textsuperscript{nd} slice)  & & &{\centering $\max_{features}=8$} \\ 
\hline
\multirow{3}{3em}{\centering 4} & \multirow{1}{7em}{\centering channel flow}  & \multirow{3}{3em}{\centering0.052}& \multirow{3}{3em}{\centering 0.086} & \multirow{1}{8em}{ \centering $n_{estimators}=50$} \\ 
& periodic hill  & & &{\centering$\max_{depth}=7$}  \\ 
&wavy wall (every 3\textsuperscript{rd} slice)  & & &{\centering $\max_{features}=8$} \\ 
\hline
\end{tabular}
    \caption[Results of the machine learning models for the converging-diverging channel test case]{\small{Results of the machine learning models for the converging-diverging channel test case.}}
    \label{tab:ml_conv_div}
\end{table}
The spatial distributed predictions and the absolute errors of every models are shown in Figure \ref{fig:ml_fall_2_mit_fehler} for cases 1 and 3. Only cases 1 and 3 are presented, since the predictions for cases 2, 3 and 4 are comparable but case 3 features the lowest error on the test data. The remarkable shape in $2.0 < x/H < 8.0$ is emerging, when data based on the wavy wall is used for training. Figure \ref{fig:feature} reveals that this is due to a certain feature, which is $q^{\left(8\right)}$ related to the eddy viscosity. Adding the wavy wall data set to the training data creates this remarkable distribution of the target quantity in a more prominent way. Overall, the results of the models are not very satisfactory. None of the models is able to correctly predict the structures of the perturbation radius in the geometry.

\noindent When training without the wavy wall, the model is able predict a greater perturbation radius for the areas at the inlet and outlet at $y/H\approx1.0$ (see areas 1 and 3 in Figure \ref{img:perRad_conv_div}). Although the model is able to recognize the distinctiveness of these areas, the absolute error reveals that the perturbation radius in far from being correctly captured. Though, the model is able to recognize these areas, which is because of the distribution of feature $q^{\left(1\right)}$, related to the Q-criterion (see Figure \ref{fig:feature}). This feature also shows greater values within these regions, which affects the prediction of the perturbation radius accordingly. However, the training data seems not to be sufficient to adequately predict the perturbation radius in some regions. Also the area above the hill at $x/H$ between $4.0$ and $8.0$ (area 2 in Figure \ref{img:perRad_conv_div}) is not correctly predicted.

To sum up, no favorable model for an application of the data-driven perturbation method could be found. Although the models are able to identify individual areas in which the perturbation radius differs from other areas, the local quantity is not predicted accurately. In order to apply the data-driven perturbation method based on a trained model, we use a model that has been trained on all data sets (case 3) subsequently. 

\begin{figure}[h!]
\centering
\includegraphics[width=0.75\textwidth]{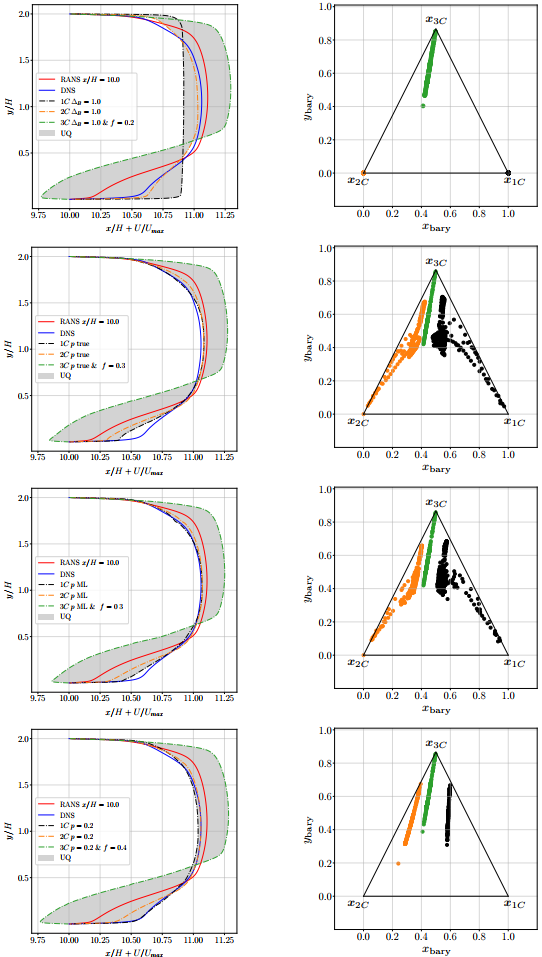}
\caption[Uncertainty bands of the converging-diverging channel calculated with different methods at $x/H=10.0$]{\small{Left: Representation of the uncertainty bands for the converging-diverging channel at $x/H=10.0$ based on four different approaches, Right: Respective perturbed barycentric coordinates.}}
\label{fig:uq_slice10}
\end{figure}

\subsubsection[Application of the data-driven perturbation method]{Application of the data-driven perturbation method}\label{abschnitt:5.4}
Since the DNS data for the converging-diverging data is entirely available in the computational domain, the correct perturbation strength (see Figure \ref{img:perRad_conv_div} can be used to determine the uncertainty estimates by the forward UQ computations. We compare the uncertainty bands based on applying the correct perturbation radius and by applying the random forest predicted perturbation radius to evaluate the potential of the machine learning method.
For the latter case, we use model 3 from Table \ref{tab:ml_conv_div}. Furthermore, we investigate how the uncertainty bands of the data-driven method change compared to the uncertainty bands of the data-free perturbation method. It can be expected that the uncertainty band of the data-driven perturbation method becomes smaller. Moreover, an additional uncertainty estimate using $p=0.2$ everywhere in the computational domain is determined in order to judge the effect of the locality of the perturbation. The value $0.2$ was chosen because it corresponds to the mean value of $p$ across our available data sets.
 
Figure \ref{fig:uq_slice10} presents the uncertainty bands at $x/H=10.0$. In addition, the position of the perturbed barycentric coordinates is presented analyze the applied perturbation strengths. Table \ref{tab:ml_uq_final} summarizes all settings and the resulting size of the enclosed area of the respective uncertainty estimates.

 \begin{table}[htb]
    \centering
    \footnotesize
    \begin{tabular}{ |c|c|c| } 
\hline
UQ method  & $f$ for $3C$-perturbation & Area of the uncertainty band at $x/H=10.0$ \\
\hline
data-free $\Delta_B=1.0$ & 0.2   & 0.910    \\
\hline
data-driven $p_{\text{true}}$ & 0.3   & 0.670   \\ 
\hline
data-driven $p_\text{ML}$ & 0.3   & 0.665   \\ 
\hline
$p=0.2$ & 0.4   & 0.758   \\

\hline
\end{tabular}
    \caption[Moderation factors and size of the uncertainty bands at $x/H=10.0$ of the converging-diverging channel]{\small{Moderation factor of the $3C$ perturbation and size of the enclosed area of the uncertainty band at $x/H=10.0$ for the applied methods.}}
    \label{tab:ml_uq_final}
\end{table}

As expected, the uncertainty bands using the data-driven method and $p=0.2$ are smaller compared to the ones resulting from the naive approach with $\Delta_B=1.0$. The uncertainty band determined using a predicted $p$ behaves very similarly to the uncertainty band determined based on the correct perturbation radius. Even if the machine learning model is not able to predict a correct $p$ locally, the uncertainty bands of the data-driven methods do not differ significantly. However, the uncertainty band established with the correct $p$ includes fewer parts of the DNS solution.
The $1C$ perturbation is even closer to the DNS solution in the case of the predicted $p$ compared to respective one based on the correct $p$. 

Moreover, no significant difference between choosing a constant value for $p$ and applying a machine learning model to locally evaluate an adequate perturbation strength can be identified. This directly leads to the fundamental question, whether the effort of the local data-driven method can be justified in the future.
 Due to the great effort in order to predict the perturbation radius locally by a machine learning model, it might make also sense to set up a method that predicts a uniform perturbation radius for a any new application.

\section{Conclusion and outlook}
The aim of this white paper was to apply and discuss the perturbation method of Emory et al. \cite{emory2013} and the data-driven extension of Heyse et al. \cite{heyse2021}. The purpose of the perturbation method is to measure the uncertainties of a RANS simulation with respect to the shape of the modeled Reynolds stress tensor.

Applying the perturbation method to the converging-diverging channel test case showed that this method is able to estimate an uncertainty bound around the original RANS solution. The larger the perturbation strength is chosen, the more the perturbed solution deviates from the original RANS simulation solution. We showed, that the choice of the corners of the barycentric triangle as limiting states of turbulence is an appropriate choice, since they lead to extreme values for QoI to a certain extend. However, especially perturbations towards the $3C$-corner can lead to convergence issues, which can be addressed by applying a moderation factor. Unfortunately, this approach seems to be impractical since the appropriate moderation factor has to be determined iteratively. We highlighted additionally, that a quantitative evaluation of the uncertainty band might be not an easy task.

Furthermore, the idea of Heyse et al. to adapt the perturbation strength locally was investigated in this work. The application of a random forest to predict the perturbation radius did not provide satisfactory results. While it is possible to learn the rough shape of the perturbation radius, even the models with the lowest errors, have not been able to correctly predict the perturbation radius locally. As expected, the uncertainty bands of the data-driven perturbation method are smaller than the uncertainty bands of the data-free approach. However, a uniform choice of $p$ lead to comparable results. Consequently, the question arises, whether training a machine learning model to predict the local perturbation strength is really worth the effort. Besides, an appropriate constant determination of perturbation strength also has to be selected based on data analysis or machine learning practices.

\end{document}